\documentclass[10pt, journal,final,compsoc]{IEEEtran}

\usepackage{graphicx}
\usepackage[section]{algorithm}
\usepackage[noend]{algorithmic}
\usepackage{amsmath}
\usepackage{bm}
\usepackage{color}
\usepackage{ragged2e}
\usepackage{subfigure}
\usepackage[numbers,sort&compress]{natbib}

\usepackage{array}
\newcommand{\PreserveBackslash}[1]{\let\temp=\\#1\let\\=\temp}
\newcolumntype{C}[1]{>{\PreserveBackslash\centering}p{#1}}
\newcolumntype{R}[1]{>{\PreserveBackslash\raggedleft}p{#1}}
\newcolumntype{L}[1]{>{\PreserveBackslash\raggedright}p{#1}}

\renewcommand{\raggedright}{\leftskip=0pt \rightskip=0pt plus 0cm}

\ifCLASSINFOpdf
\else
\fi

\begin{document}

\title{Parallel Protein Community Detection in Large-scale PPI Networks Based on Multi-source Learning}

\author{Jianguo~Chen,
        Kenli~Li, ~\IEEEmembership{Senior Member, IEEE},
        Kashif~Bilal,~\IEEEmembership{Member, IEEE},\\
        Ahmed A. Metwally,~\IEEEmembership{Member, IEEE},
        Keqin~Li, ~\IEEEmembership{Fellow, IEEE},
        and Philip S. Yu, ~\IEEEmembership{Fellow, IEEE}
\IEEEcompsocitemizethanks{\IEEEcompsocthanksitem Jianguo~Chen, Kenli~Li, and~Keqin~Li are with the College of Computer Science and Electronic Engineering, Hunan University, and the National Supercomputing Center in Changsha, Changsha, Hunan 410082, China.
\protect\\
Corresponding authors: Kenli Li and Keqin Li, Email: lkl@hnu.edu.cn, lik@newpaltz.edu.

\IEEEcompsocthanksitem Kashif Bilal is with the Comsats Institute of Information Technology, Abbottabad 45550, Pakistan.

\IEEEcompsocthanksitem  Ahmed A. Metwally is with the Departments of Bioengineering and Computer Science, University of Illinois at Chicago, IL 60607, USA.

\IEEEcompsocthanksitem Keqin Li is also with the Department of Computer Science, State University of New York, New Paltz, NY 12561, USA.

\IEEEcompsocthanksitem Philip S. Yu is with the Department of Computer Science, University of Illinois at Chicago, Chicago, IL 60607, USA.
}
}

\markboth{}{Shell \MakeLowercase{\textit{et al.}}: Bare Advanced Demo of IEEEtran.cls for Journals}

\IEEEtitleabstractindextext{
\renewcommand{\raggedright}{\leftskip=0pt \rightskip=0pt plus 0cm}
\begin{abstract}
 \raggedright{
Protein interactions constitute the fundamental building block of almost every life activity.
Identifying protein communities from Protein-Protein Interaction (PPI) networks is essential to understand the principles of cellular organization and explore the causes of various diseases.
It is critical to integrate multiple data resources to identify reliable protein communities that have biological significance and improve the performance of community detection methods for large-scale PPI networks.
In this paper, we propose a Multi-source Learning based Protein Community Detection (MLPCD) algorithm by integrating Gene Expression Data (GED) and a parallel solution of MLPCD using cloud computing technology.
To effectively discover the biological functions of proteins that participating in different cellular processes,
GED under different conditions is integrated with the original PPI network to reconstruct a Weighted-PPI (WPPI) network.
To flexibly identify protein communities of different scales, we define community modularity and functional cohesion measurements and detect protein communities from WPPI using an agglomerative method.
In addition, we respectively compare the detected communities with known protein complexes and evaluate the functional enrichment of protein function modules using Gene Ontology annotations.
Moreover, we implement a parallel version of the MLPCD algorithm on the Apache Spark platform to enhance the performance of the algorithm for large-scale realistic PPI networks.
Extensive experimental results indicate the superiority and notable advantages of the MLPCD algorithm over the relevant algorithms in terms of accuracy and performance.
}
\end{abstract}

}

\maketitle
\IEEEdisplaynontitleabstractindextext
\IEEEpeerreviewmaketitle

\section{Introduction}
\subsection{Motivation}
\IEEEPARstart{T}{aking} advantage of high-throughput technologies, massive Protein-Protein Interaction (PPI) datasets and other biological substance data are accumulated \cite{a01}.
PPI networks contain valuable knowledge, which can further aid to understand the principles of cellular organization, identify various disease mechanisms, and serve as a basis for new therapeutic approaches \cite{a02}.
Various research efforts indicate that proteins have a tendency to perform one or more cell activities via interacted proteins \cite{a01, a03}.
Abundant clustering and community detection methods have been proposed to discover protein communities and modules from PPI networks, such as the Girvan and Newman (GN) \cite{a04}, Louvain \cite{a05}, Label Propagation (LP) \cite{a06}, and SPICi \cite{a07} algorithms.
Accurate protein communities in PPI datasets are useful for downstream analysis, such as the identification of the potential protein complexes and functional modules, as well as the understanding of the molecular machines and metabolism pathways \cite{a09}.

While significant efforts have been contributed to the PPI data analysis, there are two major challenges in protein community detection of PPI networks.
Firstly, traditional approaches generally identify protein communities only according to the topological structure of PPI networks, where the identified modules may not effectively reflect their biological significance \cite{a10}.
In fact, proteins in a PPI network are intrinsically controlled by different regulatory mechanisms in different biological processes, appearing in different functions \cite{a11}.
Studies on gene expression indicate that the similarity in biological role often corresponds to high gene co-expression \cite{a12, a13}.
Secondly, the performance improvement of protein community detection in large-scale PPI networks is also significant.
Especially in the era of Broad Learning (BL) \cite{a14}, it is essential to efficiently integrate multiple data sources and quickly discover knowledge from different data sources.
Stand-alone and serial algorithms are inefficient and unable to scale to the exponentially increasing complexity of PPI network data.

\subsection{Our Contributions}
In this paper, we aim to enhance efficiency and accuracy of protein community detection of large-scale PPI networks.
We propose a model to integrate the original PPI network and the related Gene Expression Data (GED) to form a Weighted PPI (WPPI) network, where the gene co-expression in a specific biological process is considered.
We propose a Multi-source Learning based Protein Community Detection (MLPCD) algorithm to identify protein communities on the WPPI networks in an agglomerative way.
We parallelize MLPCD on a cloud computing platform to process large-scale WPPI networks.
Extensive experimental results indicate the MLPCD algorithm achieves high accuracy and performance in comparison with other algorithms.
Our contributions in this paper are summarized as follows.
\begin{itemize}
  \item We reconstruct a WPPI network by integrating the original PPI network and the related GED dataset.
        The topological structure of proteins in PPI networks and their gene correlation in different biological processes are considered.
  \item We define community modularity and functional cohesion measurements to flexibly identify protein communities of different scales with the corresponding biological significance.
        Then, an agglomerative Louvain method is introduced to detect protein communities from the WPPI network.
  \item To improve the performance of MLPCD algorithm and efficiently detect protein communities from large-scale WPPI networks, we propose a parallel solution of MLPCD using the Apache Spark cloud platform.
  \item We evaluate the detected protein communities in terms of protein complexes and functional modules, comparing the detected communities with known protein complexes and evaluate the functional enrichment of protein function modules using Gene Ontology annotations.
\end{itemize}

The remainder of the paper is structured as follows.
Section 2 reviews the related work.
Section 3 presents the construction of the WPPI network and MLPCD algorithm.
Parallel implementation of the MLPCD algorithm is described in detail in Section 4.
Experimental evaluations of the proposed algorithm are presented in Section 5.
Section 6 concludes the paper with future work and directions.

\section{Related Work}
The analysis of protein functions in PPI networks is a hot topic in Bioinformatics research.
The main research streaming for PPI networks focuses on the detection of protein functional modules and analysis of the dynamic PPI networks' characteristic \cite{a02}.
Peng \emph{et al}. \cite{a12} proposed a Udonc algorithm for identifying essential proteins based on protein domains and PPI networks.
Zhu \emph{et al}. \cite{a15} introduced a local similarity preserving embedding algorithm to identify spurious interactions in PPI networks.
Sanghamitra \emph{et al}. \cite{a16} proposed a new feature vector based on gene ontology terms for PPI prediction, in which a protein pair is considered as a document and the terms annotating the two proteins represent the words.
Ji \emph{et al}. \cite{a10} compared some functional module detection methods for PPI, and discussed the accuracy and performance of several typical algorithms.
Li \emph{et al}. \cite{a11} discussed a topology potential-based method for identifying essential proteins from PPI networks.
The previous protein modules detection approaches of PPI networks achieved a certain degree of success.
However, most of the existing achievements detect protein communities only according to the topological structure of PPI networks, where the identified protein sub-cluster may not effectively reflect its biological significance.

Existing studies demonstrate that the genes or gene products with similar expression patterns tend to have the similar biological function in a period of life, and also more likely to contact each other to form a dense functional module in PPI networks \cite{a17}.
Therefore, proteins' GED data are used in this work to evaluate the similarity of proteins in a PPI network \cite{a18,a09}.
Ji \emph{et al}. \cite{a18} introduced a multiple-grain model to detect functional modules from large-scale PPI networks.
Spirin \emph{et al}. \cite{a09} presented an enumeration method to find completely connected subgraphs, and then to search for protein functional module.
GN algorithm \cite{a04} is a classical community discovery algorithm.
Depending on high cohesion for internal communities and low cohesion among communities, structures of cohesive communities are relatively detected by gradually removing the edges among communities.
Louvain \cite{a05} is a fast aggregation algorithm that is used to extract the community structure in large networks, using a heuristic method based on modularity optimization.
SPICi \cite{a07} is a clustering algorithm for discovering protein complexes and functional modules from PPI networks.

Focusing on the performance of protein community detection algorithms for large-scale PPI data, numerous studies on the intersection of parallel/distributed computing and the community detection models were proposed.
Krejci \emph{et al}. \cite{a19} introduced a hidden Markov model-based peptide clustering algorithm to identify protein-interaction consensus motifs in large datasets.
Bader \emph{et al}. \cite{a20} proposed a graph-theoretic analysis of the human protein-interaction network using multi-core parallel algorithms.
Many state-of-the-art technologies, such as cloud and distributed computing offer high-speed computing power.
In \cite{a21, a22}, various algorithms were proposed based on the MapReduce model of Apache Hadoop.
Apache Spark \cite{a23} is an excellent cloud platform that is suitable for data mining.
The Spark platform saves huge amounts of disk I/O operation, and is more suitable for data mining with iterative computation.
Therefore, to handle large-scale PPI networks, we propose a parallel algorithm to efficiently detect protein communities based on the Apache Spark cloud environment.

\section{Multi-source Learning Based Protein Community Detection Algorithm}
\subsection{MLPCD Architecture}
\begin{figure}[!ht]
\setlength{\abovecaptionskip}{0pt}
\setlength{\belowcaptionskip}{0pt}
\centering
\includegraphics[width=3.4in]{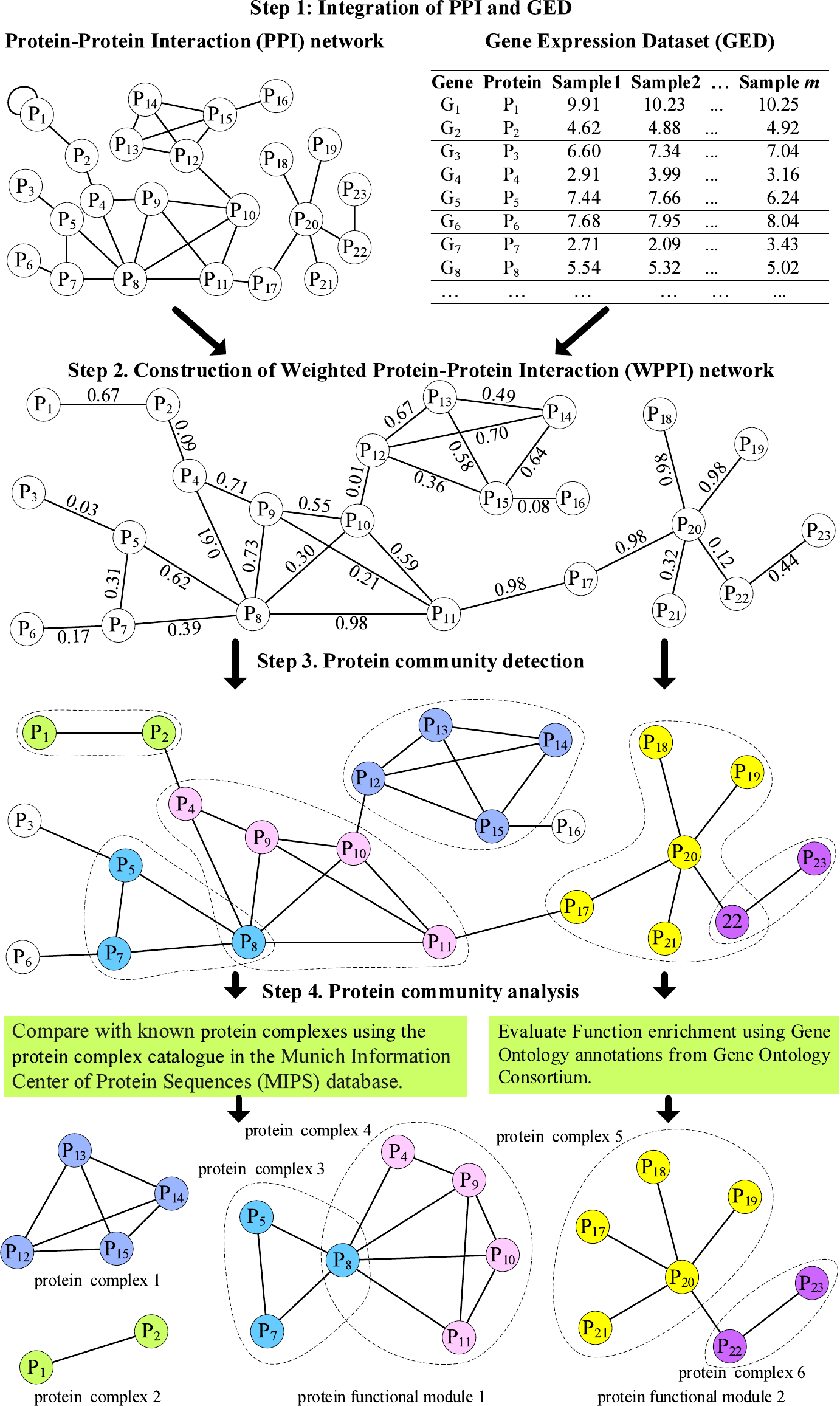}
\caption{Example of the architecture of the proposed MLPCD algorithm. MLPCD consists of four steps. Step 1 is described in Section \ref{section3.2.1}, Step 2 is described in Section \ref{section3.2.2}, Step 3 is presented in Section \ref{section3.3}, and Step 4 is given in Section \ref{section5.2}.}
\label{fig01}
\end{figure}

In this section, we propose a MLPCD algorithm to detect protein communities from large-scale PPI networks by integrating information from the GED.
A Weighted PPI network is built based on the original PPI network and the related GED datasets in specific biological processes.
We define community modularity and functional cohesion measurements to flexibly identify protein communities of different scales with the corresponding biological significance.
In addition, an agglomerative Louvain method is introduced to detect protein communities from the WPPI network.
An example of the architecture of the proposed MLPCD algorithm is illustrated in Fig. \ref{fig01}.

\subsection{Weighted PPI Network Construction Based on GED}

\subsubsection{Integration of PPI and GED}
\label{section3.2.1}
Although a gene expression level cannot always represent its protein concentration, previous studies \cite{a13} have observed notable correlations between them.
We estimate the gene co-expression of proteins in a PPI network.
Proteins with similarly co-expression genes are more likely to be linked with each other in a PPI network to form a dense functional module.
Thus, GED is introduced to investigate the co-expression of protein interaction.
A PPI network is represented as an undirected graph $G_{PPI}(V,~E)$, where $V$ is a collection of protein nodes in the graph,
and $E$ represents a collection of edges among the nodes.
The interactive relationship among proteins in a PPI network is described as an adjacency matrix $A$, as defined in Eq. (\ref{eq01}):
\begin{equation}
\setlength{\abovecaptionskip}{0pt}
\setlength{\belowcaptionskip}{0pt}
\label{eq01}
A =
\begin{bmatrix}
a_{11}  & \dots & a_{1N}  \\
\vdots  & \ddots & \vdots\\
a_{N1}  & \dots & a_{NN} \\
\end{bmatrix},
\end{equation}
where $N$ is the number of protein nodes in the PPI network, and $a_{ij} = 1$ if there exists an interactive relationship between proteins $x_{i}$ and $x_{j}$, otherwise, $a_{ij} = 0$.

To evaluate the gene co-expression of proteins in a PPI network, we collect GED datasets in specific conditions for each protein.
Assume that there are $N$ genes required to be compared, and $M$ microarray experiments for each gene, the set of $GED$ will be obtained from these experiments, as defined as:

\begin{equation}
\setlength{\abovecaptionskip}{0pt}
\setlength{\belowcaptionskip}{0pt}
\label{eq02}
GED =
\begin{bmatrix}
G_{1}\\
\vdots\\
G_{N}\\
\end{bmatrix}
=
\begin{bmatrix}
g_{11} & \dots & g_{1M} \\
\vdots & \ddots & \vdots\\
g_{N1} & \dots & g_{NM} \\
\end{bmatrix}
,
\end{equation}
where $N$ is the number of genes in a $GED$ dataset in a specific condition, and $M$ is the number of microarray experiments.
$G_{i} = \{g_{i1}, ~g_{i2}, ~\dots, ~g_{iM}\}$.
Considering the differential expression for microarray data, we use the Limma quantile normalization \cite{a24} for analysis the GED from microarray experiments.

For each two-proteins-pair of an edge in a PPI network, we compute the co-expression of two corresponding genes.
Gene co-expression is measured by the similarity of the corresponding gene expression data.
Because genes expression from microarray experiments hold the characteristic of continuous, normal distribution, and linearly related \cite{a13}, we use the Pearson correlation coefficient method to calculate the similarity of genes.
The Pearson correlation coefficient of $G_{i}$ and $G_{j}$ is defined in Eq. (\ref{eq03}):

\begin{equation}
\setlength{\abovecaptionskip}{0pt}
\setlength{\belowcaptionskip}{0pt}
\label{eq03}
p_{ij}= P(G_{i}, G_{j}) =
\frac{1}{M} \sum_{k=1}^{M}{\frac{g_{ik}-\overline{G_{i}}}{\sigma(G_{i})} \times \frac{g_{jk}-\overline{G_{j}}}{\sigma(G_{j})}},
\end{equation}
where $M$ is the number of samples in a condition-specific experiment, $g_{ik}$ and $g_{jk}$ are the expression levels of genes $G_{i}$ and $G_{j}$ in the $k$-th sample, $\overline{G_{i}}$ is the average expression level value of gene $G_{i}$ in $M$ experiments, and $\sigma(G_{i})$ is the standard deviation of the gene expression vectors of gene $G_{i}$.
The value of $p_{ij}$ is in the range of $[-1, ~1]$.
The two genes $G_{i}$ and $G_{j}$ are correlated when $p_{ij} = 1$, unrelated when $p_{ij} =0$, and anti correlated when $p_{ij} = -1$.
The larger the Pearson correlation coefficient is, the more similar the corresponding two genes are.

\subsubsection{Construction of Weighted PPI Network}
\label{section3.2.2}
We compute the similarity of all protein entries in the original PPI network, and combine the entries-similarity of proteins and the interactions among these proteins.
Considering that the inverse expression profiles of proteins in the same community, we assign a non-negative weight to each interaction.
The weighted gene co-expression of two proteins $x_{i}$ and $x_{j}$ is defined in Eq. (\ref{eq04}):

\begin{equation}
\label{eq04}
a_{ij}^{w} = |p_{ij}| \times a_{ij}.
\end{equation}
For each edge $e_{ij}$ between proteins $x_{i}$ and $x_{j}$, the absolute value of their gene co-expression is used as the edge weight.
The interactive relationship among proteins in a PPI network is re-described as a weighted adjacency matrix $A^{W}$, as defined in Eq. (\ref{eq05}):

\begin{equation}
\label{eq05}
A^{W} =
\begin{bmatrix}
a_{11}^{w}  & \dots   & a_{1N}^{w}  \\
\vdots      & \ddots  & \vdots\\
a_{N1}^{w}  & \dots   & a_{NN}^{w}  \\
\end{bmatrix}.
\end{equation}

On the base of the weighted adjacency matrix, a WPPI network is constructed.
We propose the definition of a WPPI network as follows.

\textbf{Definition 1: Weighted PPI (WPPI) network}.
\textit{Given a PPI network $G_{PPI}(V, E)$, the related gene expression dataset is collected to calculate the gene co-expression among proteins.
A weighted adjacency matrix $A^{W}$ is created by integrating the gene co-expression and the interactive relationships of proteins.
A WPPI network $G_{WPPI}(V,E^{W})$ is built based on the weighted adjacency matrix, where each edge $e_{ij}$ $\in$ $E^{W}$ holds a weighted value $a_{ij}^{w}$.}

The detailed steps of the WPPI network construction are presented in Algorithm \ref{alg31}.
Assume that there are $m$ edges in a PPI network, the time complexity of the construction process of a WPPI network is $O(mM)$, where $M$ is the number of samples of each GED record.

\begin{algorithm}[!ht]
\scriptsize
\caption{{\scriptsize Construction process of WPPI networks}}
\label{alg31}
\begin{algorithmic}[1]
\REQUIRE ~\\
    $PPI$: the dataset of a PPI network;\\
    $GED$: the gene expression data of all proteins in PPI.\\
\ENSURE ~\\
    $WPPI$: the weighted PPI network.
\FOR {each edge $e$ in $E(PPI)$}
    \STATE protein-pair $(x_{i}, ~x_{j})$ $\leftarrow$ $e$;
    \STATE calculate the adjacency matrix $a_{ij}$ of $(x_{i}, ~x_{j})$;\\
    \STATE $(G_{i}, ~G_{j})$ $\leftarrow$ loadGenes($x_{i}, ~x_{j}, ~GED$);
    \FOR {each gene in $(G_{i}, ~G_{j})$}
    \STATE calculate Pearson correlation coefficient $p_{ij} \leftarrow
\frac{1}{M} \sum_{k=1}^{M}{\frac{g_{ik}-\overline{G_{i}}}{\sigma(G_{i})} \times \frac{g_{jk}-\overline{G_{j}}}{\sigma(G_{j})}}$;
    \ENDFOR
\ENDFOR
\FOR {each protein-pair ($x_{i},~x_{j}$) in $PPI$}
\STATE $a_{ij}^{w} \leftarrow |p_{ij}| \times a_{ij}$;
\STATE $WPPI$.$A^{W}$($a_{ij}^{w}$)
\ENDFOR
\RETURN $WPPI$.
\end{algorithmic}
\end{algorithm}

\subsection{MLPCD Algorithm for WPPI Networks}
\label{section3.3}
On the basis of the WPPI network, we propose a MLPCD algorithm to detect protein communities from WPPI network effectively and make protein communities more biologically significant.
The MLPCD algorithm is designed based on the Louvain algorithm \cite{a05}.
Similar to Louvain, we first initialize the communities based on vertices and then calculate the community modularity to generate communities in an agglomerative way.
Different from Louvain, we optimize the original Louvain algorithm for protein community detection in three aspects:
(1) We calculate the weighted degree of each vertex and select the vertices with high weighted degrees as the initial communities.
(2) We define a new modularity for protein communities by considering the characteristics of protein modules, such as protein complex and functional modules.
(3) A functional cohesion measurement is proposed to evaluate the detected protein communities, making the detected protein communities to have more biological significance.

\subsubsection{Community Initialization}
Given a WPPI network $G_{WPPI}(V, E^{W})$, where $V$ is a set of protein entities, and $E^{W}$ is a set of interactive relationships among these proteins with the related gene co-expression.
We calculate the weighted degree for each vertex.
For a vertex $v_{i}$, the vertices connected to $v_{i}$ are defined as its neighbors $N(v_{i})$.
The weighted degree of a vertex $v_{i}$ is donated as $d(v_{i})$, which is the sum of the weight values of edges connecting $v_{i}$, as defined in Eq. (\ref{eq06}):

\begin{equation}
\label{eq06}
d(v_{i}) = \sum_{e_{ij} \in E^{W}}{a_{ij}^{w}},
\end{equation}
where $a_{ij}^{w}$ is the weight value of $e_{ij}$ between vertices $v_{i}$ and $v_{j}$.
Since the weight value of each edge is measured by the related gene co-expression, the weighted degree $d(v_{i})$ can reflect the biological activity of the vertex $v_{i}$ and its neighbors $N(v_{i})$.

By analyzing the structures of known protein functional modules of PPI networks, we find that there exists one type of proteins (termed as ``hub proteins'') that has frequent interactions with their neighbors.
Similar phenomena have been confirmed in the previous studies, where the hub proteins are evaluated that play a dominant role in maintaining the functionality of PPI networks \cite{a26}.
Therefore, we define the hub vertices of a WPPI network based on the weighted degree of vertices.
The hub vertices are selected by higher than a specified threshold $d_{\alpha}$.
Experimental results show that the threshold $d_{\alpha}$ finds the best effectiveness at the value of $d_{\alpha} = \frac{1}{N}\sum_{i=1}^{N}{d(v_{i})}$.
In this way, these hub vertices are used as initial communities.

\subsubsection{Community Modularity}

We concentrate on two types of protein communities: protein complexes and protein functional modules.
Protein complexes are sets of proteins that interact with each other to execute a single multimolecular mechanism \cite{a09}.
Protein functional modules are sets of proteins that participate in a particular biological process, interacting with each other at different time and places.
According to the characteristics of protein complexes and protein functional modules, we define the protein community as a group of proteins that share genes or cellular interactions among them, and are separable from those of other communities.
The modularity $Q_{C_{k}}$ of a protein community $C_{k}$ in a WPPI network is defined as the ratio of in-degrees and out-degrees of proteins in $C_{k}$, as defined in Eq. (\ref{eq07}).

\begin{equation}
\label{eq07}
Q_{C_{k}} = \frac{\sum{d_{i}^{in}(C_{k},v_{i})}}{\sum{d_{i}^{out}(C_{k},v_{i})}}, ~~~~\forall {v_{i} \in C_{k}},
\end{equation}
where $d_{i}^{in}(C_{k},v_{i})$ is the in-degree of the vertex $v_{i}$ in $C_{k}$, and $d_{i}^{in}(C_{k},v_{i}) = \sum_{v_{j} \in C_{k}}{a_{ij}^{w}}$.
$d_{i}^{out}(C_{k},v_{i})$ is the out-degree of the vertex $v_{i}$ in $C_{k}$, and $d_{i}^{in}(C_{k},v_{i}) = \sum_{v_{j} \notin C_{k}}{a_{ij}^{w}}$.

For each hub protein vertex $v_{i}$ obtained in the community initialization stage, we collect its neighbors $N(v_{i})$ and try to append each neighbor $v_{j} \in N(v_{i})$ to the community of $v_{i}$.
Then, we evaluate the updating modularity $\Delta{Q_{C_{k}}}$ that caused by $v_{j}$, as defined in Eq. (\ref{eq08}):

\begin{equation}
\label{eq08}
\Delta{Q_{C_{k},v_{j}}} = Q_{C_{k} \cup v_{j}} - Q_{C_{k}}.
\end{equation}
We respectively calculate the updating modularity $\Delta{Q_{C_{k},v_{j}}}$ for each neighbor in $N(v_{i})$ and record the neighbor vertex with the maximum $\Delta{Q_{C_{k},v_{j}}}$.
If $max \Delta{Q_{C_{k},v_{j}}}>0$, the related vertex is appended to the community $C_{k}$ of $v_{i}$.
Then, for each vertex in the updated community $C_{k}$, we collect its neighbors and evaluate their contribution to the modularity $\Delta{Q}$.
Repeat this process, until the modularity of all protein communities is stable.
According to the modularity of protein communities, most of the protein complexes and parts of protein functional modules are detected effectively.

\subsubsection{Functional Cohesion}

In the second stage of the MLPCD algorithm, the communities detected in the first stage are compressed to construct a new network $G'(V',E')$.
Each new vertex $v_{i}'$ in $V'$ is created from the related community $C_{i}$, and the weighted degree $d(v_{i}')$ of $v_{i}'$ is calculated by the total value of weighted degrees of all vertices in $C_{i}$, as defined in Eq. (\ref{eq09}):
\begin{equation}
\label{eq09}
d(v_{i}') = \sum_{v_{j} \in C_{i}}{d(v_{j})}.
\end{equation}
The weight value $a_{ij}^{w'}$ of edge $e_{ij}'$ between vertices $v_{i}'$ and $v_{j}'$ is defined by the total weight values of edges among vertices in $C_{i}'$ and $C_{j}'$, as defined in Eq. (\ref{eq10}):
\begin{equation}
\label{eq10}
a_{ij}^{w'} = \sum_{e_{ij}' \in E'}{a_{ij}^{w}}, \forall v_{i} \in C_{i}, v_{j} \in C_{j}.
\end{equation}

For the new network $G'(V',E')$, we initial each new vertex $v_{i}'$ in $G'$ as a community $C_{i}'$.
Then, we gradually append the neighbors of $v_{i}'$ and evaluate their contribution in terms of connectivity, interaction intensity, and functional cohesion.
Connectivity is a key characteristic of communities in a network, which is defined as the ratio of the number of edges among the vertices in the community and the maximum possible number of edges among them.
The connectivity $Con_{k}$ of a community $C_{k}'$ is calculated in Eq. (\ref{eq11}):
\begin{equation}
\label{eq11}
Con_{k} = \frac{2|E_{k}|}{|V_{k}|\times(|V_{k}|-1)},
\end{equation}
where $|E_{k}|$ is the number of existing edges in $C_{k}'$ and $|V_{k}|$ is the number of vertices in $C_{k}'$.

For protein communities in WPPI networks, besides connectivity, interaction intensity is another important feature.
Given a set of vertices for a candidate community, the weighted degrees of these vertices are calculated, respectively.
Then, the interaction intensity of the candidate community is defined as the ratio of the in-degree of this community and the sum of the average weighted-degrees of all vertices in it, as calculated in Eq. (\ref{eq12}):
\begin{equation}
\label{eq12}
II_{k} = \frac{2\sum\limits_{v_{i}',v_{j}' \in C_{k}'}{a_{ij}^{w'}}}{\sum\limits_{v_{i}' \in C_{k}'}{\overline{d(v_{i}')}}},
\end{equation}
where $\overline{d(v_{i}')} = \frac{1}{|N(v_{i}')|}\sum_{v_{j}' \in N(v_{i}')}{a_{ij}^{w'}}$.
Intuitively, given a vertex and all its neighbors, we divide these neighbors into high-quality neighbors and low-quality neighbors based on the edge weights between them.
Then, in a community, if each vertex has more high-quality neighbors in the same community, the interaction intensity of the community is higher.
Based on the connectivity and interaction intensity of protein communities, we propose a definition of Functional Cohesion (FC) for protein communities.

\textbf{Definition 2: Functional Cohesion.}
\textit{The functional cohesion of a protein community reflects the connectivity and interaction intensity of all the proteins in the community.
A protein community with high functional cohesion requires not only a high intensity of interaction among proteins, but also a high degree of connectivity.
The value of functional cohesion of a protein community is calculated by the product of the values of interaction intensity and connectivity.}

\begin{figure}[!ht]
\setlength{\abovecaptionskip}{0pt}
\setlength{\belowcaptionskip}{0pt}
\centering
\subfigure[Two-vertices community]{
 \label{fig02:a}
 \includegraphics[width=1.5in]{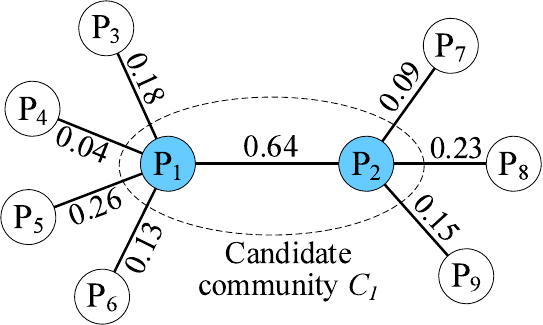}}
 \subfigure[Six-vertices community]{
 \label{fig02:b}
 \includegraphics[width=1.5in]{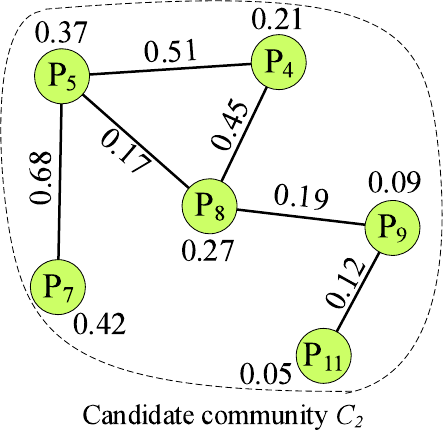}}
\caption{Examples of functional cohesion of protein communities.}
\label{fig02}
\end{figure}

Given a candidate community $C_{k}$, according to the interaction intensity $II_{k}$ and connectivity $Con_{k}$, the functional cohesion $FC_{k}$ is calculated in Eq. (\ref{eq13}):

\begin{equation}
\label{eq13}
\begin{aligned}
FC_{k} & = II_{k} \times Con_{k} \\
          & = \frac{2\sum\limits_{v_{i}',v_{j}' \in C_{k}}{a_{ij}^{w'}}}{\sum\limits_{v_{i}' \in C_{k}}{\overline{d(v_{i}')}}} \times \frac{2|E_{k}|}{|V_{k}|\times(|V_{k}|-1)}.
\end{aligned}
\end{equation}
Examples of functional cohesion of protein communities are illustrated in Fig. \ref{fig02}.
In Fig. \ref{fig02:a}, there are two vertices $P_{1}$ and $P_{2}$ in the candidate community $C_{1}$, we can easily calculate the connectivity $Con_{1} = 1$ and the interaction intensity $II_{1} = 2.43$.
Therefore, the functional cohesion of $C_{1}$ is obtained as $FC_{1} = 2.43$.
In Fig. \ref{fig02:b}, there are six vertices in the candidate community $C_{2}$ with six edges.
Considering that the maximum possible number of edges in $C_{2}$ is equal to $30$, we can calculate the connectivity $Con_{2} = 12/30 = 0.4$.
Based on the given average weighted-degree of each vertex, we calculate the interaction intensity $II_{2} = \frac{2.12 \times 2}{1.41} = 3.01$.
Hence, the functional cohesion of $C_{2}$ is obtained as $FC_{2} = 1.21$.

To obtain protein communities with flexible scales, we define a parameter $\lambda$ as the threshold of the functional cohesion.
A set of proteins forms a community if their functional cohesion exceeds $\lambda$.
Experimental results show that $\lambda$ finds the best effectiveness in the range of (1.0 - 3.0).
Repeat this process, until all vertices that satisfy $\lambda$ are assigned to the related communities.
The larger the $FC$ function value of a community, the closer it will be to the actual community structure of the network.
The description of the MLPCD algorithm for protein community detection is presented in Algorithm \ref{alg32}.

\begin{algorithm}[!ht]
\scriptsize
\caption{{\scriptsize MLPCD algorithm}}
\label{alg32}
\begin{algorithmic}[1]
\REQUIRE ~\\
    $WPPI$: the dataset of a WPPI network;\\
    $d_{\alpha}$: the threshold of the initial hub vertices; \\
    $\lambda$: the threshold of functional cohesion.
\ENSURE ~\\
    $PCs$: the detected protein communities.
\FOR {vertex $v_{i}$ in WPPI}
\STATE calculate the weighted degree $d(v_{i})$;
\IF {$d(v_{i})$ $>$ $d_{\alpha}$}
\STATE select $v_{i}$ as initial community $\rightarrow Cs$;
\ENDIF
\ENDFOR
\FOR {each community $C_{k}$ in $PCs$}
\STATE collect neighbors $N(C_{k})$ of vertices in $C_{k}$;
\FOR {each vertex $v_{j}$ in $N(C_{k})$}
\STATE calculate the modularity $\Delta{Q_{C_{k},v_{j}}} \leftarrow Q_{C_{k} \cup v_{j}} - Q_{C_{k}}$;
\ENDFOR
\STATE append the neighbor with max $\Delta{Q}$ to $C_{k}$;
\ENDFOR
\STATE reconstruct new network $G'$ based on communities $Cs$;
\STATE initial vertices as communities;
\FOR {vertex $v_{i}'$ in $G'$}
\STATE collect neighbors $N(v_{i}')$ of $v_{i}'$;
\FOR {each vertex $v_{j}'$ in $N(v_{i}')$}
\STATE append $v_{j}'$ to the community $C_{k}'$ that $v_{i}'$ is located;
\STATE calculate functional cohesion $FC_{k} \leftarrow II_{k} \times Con_{k}$;
\IF {$FC_{k} < \lambda$}
\STATE remove $v_{j}'$ from the community $C_{k}'$;
\ENDIF
\ENDFOR
\ENDFOR
\RETURN $PCs$.
\end{algorithmic}
\end{algorithm}

Assume that there are $N$ proteins in the WPPI network, the number of communities detected in the first stage is equal to $K_{1}$, the time complexity of community initialization is $O(N)$.
The time complexity of the first and second stages is $O(NK_{1})$ and $O(K_{1})$, respectively.
Therefore, the time complexity of the MLPCD algorithm is $O(NK_{1})$.

\section{Parallelization for MLPCD Algorithm}
To improve the performance of the proposed MLPCD algorithm and efficiently handle large-scale PPI networks, we propose a parallelization solution for MLPCD on the Apache Spark cloud computing platform.
The processes of WPPI network construction and protein community detection are executed in parallel, respectively.

\subsection{Parallel Construction of WPPI Networks}
In the parallel process of WPPI network construction, RDD objects of the original PPI network and GED are created, which consists of multiple partitions that support parallel computing.
Then, logical and data dependencies among these RDD objects and partitions are analyzed according to the processes of gene co-expression and weighted adjacency matrix calculation.
The terminology acronyms and functions of the parallel solution are described in Tables \ref{table41} and \ref{table42}.

\begin{table}[!ht]
\scriptsize
\renewcommand{\arraystretch}{1.0}
\caption{{\scriptsize Terminology acronyms of Spark-based parallel implementation.}}
\label{table41}
\centering
\tabcolsep1pt
\begin{tabular}{L{1.5cm} L{7.0cm}}
\hline
Acronyms &  Description\\
\hline
RDD     &the resilient distributed datasets supported by Apache Spark, which is a resilient and distributed collection of records spread over one or many partitions.\\
$RDD_{PPI}$ & the RDD object of the original PPI network. \\
$RDD_{GED}$ & the RDD object of the GED dataset corresponding to $RDD_{PPI}$.\\
$RDD_{P}$ & the RDD object of the results of Pearson correlation coeffcient.\\
$RDD_{A}$ & the adjacency matrix of the original PPI network, where each element refers the edge between two proteins. \\
$RDD_{A^{W}}$ & the weighted adjacency matrix obtained based on $RDD_{P}$ and $RDD_{A}$. \\
$RDD_{WPPI}$ &the RDD object of the weighted PPI network.\\
\hline
\end{tabular}
\end{table}

\begin{table}[!ht]
\scriptsize
\renewcommand{\arraystretch}{1.0}
\caption{{\scriptsize RDD-based functions supported by the Apache Spark.}}
\label{table42}
\centering
\tabcolsep1pt
\begin{tabular}{L{1.6cm} L{6.9cm}}
\hline
Functions & Description\\
\hline
map()  & Each partition in an RDD object is calculated in parallel by a user-defined function and transformed to a new RDD.\\
flatMap() & Similar to map(), but each partition can be mapped to zero or more partitions, and finally be flatted and output.\\
reduceBykey() &Merge the values of RDD objects with each key.\\
reduce() & Calculate two elements of the input RDD at one time by a user-defined function and then generates a new element, which will be calculated with the next element.\\
cache()  & Save the output RDD object to the main memory, which will be used in the next process. \\
\hline
\end{tabular}
\end{table}

\subsubsection{RDD Object Dependence}
\begin{figure}[!ht]
\setlength{\abovecaptionskip}{0pt}
\setlength{\belowcaptionskip}{0pt}
\centering
\includegraphics[width=3.0in]{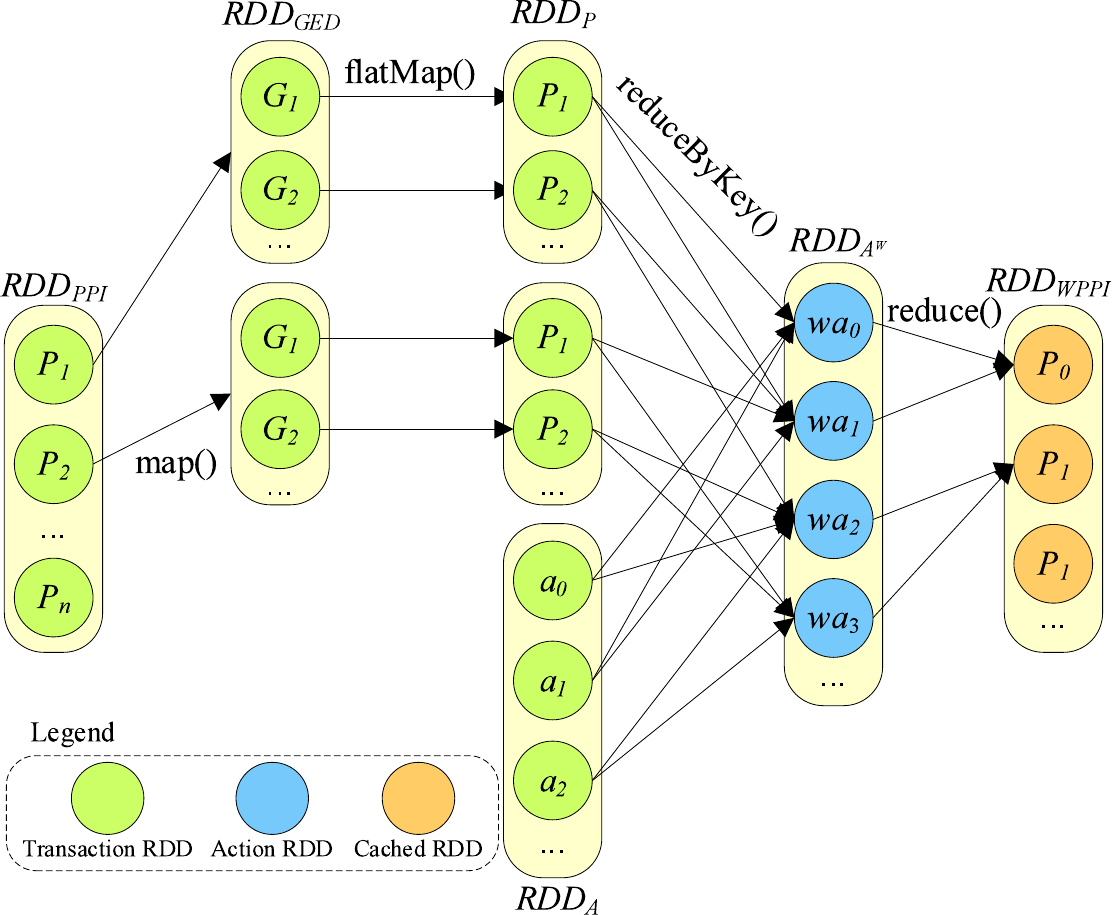}
\caption{RDD dependence of WPP parallel construction.}
\label{fig03}
\end{figure}

At the beginning of the construction process of the WPPI network, the datasets of the original PPI network and GED are loaded into the Spark Tachyon memory system as an RDD object $RDD_{PPI}$.
In Spark, each RDD object is an expressive form of a dataset in a definite state, which might be transformed from a prior state.
In other words, there might be a dependence between the current RDD object and the prior RDD(s).
RDD object dependence of the WPPI network construction is illustrated in Fig. \ref{fig03}.

In Fig. \ref{fig03}, an RDD object $RDD_{PPI}$ is established for the PPI network, which consists multiple partitions $P_{i}$.
The corresponding GED dataset of each partition is loaded as $RDD_{GED}$.
There exist narrow dependencies between $RDD_{GED}$ and $RDD_{PPI}$.
In a narrow dependence, each partition of an RDD is utilized by no more than one partition of the child RDD.
We calculate the Pearson correlation coefficient of genes in each $RDD_{GED}$ in parallel, and obtain the results of gene co-expression $RDD_{P}$ in the $flatMap()$ function.
Then, $RDD_{P}$ is integrated with the adjacency matrix $RDD_{A}$ to generate the weighted adjacency object $RDD_{A^{W}}$.
Hence, there exists a wide dependence between $RDD_{A^{W}}$ and $RDD_{A}$, $RDD_{P}$, respectively.
A wide dependence refers to each partition of the sub-RDD depends on multiple partitions of the parent RDD, that is, there is a partition of the parent RDD corresponding to multiple partitions of the sub-RDD.
Finally, the RDD object $RDD_{WPPI}$ of the WPPI network is obtained depending on $RDD_{A^{W}}$ in the $reduce()$ function, and is cached in the Tachyon system to be utilized in the coming protein community detection process.

\subsubsection{Parallel Execution Process of WPPI Construction}
Based on the RDD dependence of the WPPI network construction process, a task DAG is built to generate parallel jobs and tasks of WPPI network construction.
The construction job of the WPPI network is submitted from a driver computer to the master computer of the Spark cluster.
A job scheduling module DAGScheduler of Spark is available for the submitted construction job.
DAGScheduler analyzes the submitted job and divides it into multiple job stages according to the RDD dependence.

As shown in Fig. \ref{fig03}, there are 3 stages in the WPPI network construction job.
In stage 1, $RDD_{P}$ is created from $RDD_{GED}$ in the $map()$ and $flatMap()$ functions, with narrow dependencies among them.
In stage 2, $RDD_{A^{W}}$ is calculated from $RDD_{P}$ and $RDD_{A}$ with wide dependencies.
In stage 3, $RDD_{WPPI}$ is obtained from $RDD_{A^{W}}$ with a narrow dependence.
Each job stage is further divided into multiple tasks, which will be allocated to different computers and executed in parallel.
The detail steps of the parallel construction process are described in Algorithm \ref{alg401}.

\begin{algorithm}[!ht]
\scriptsize
\caption{{\scriptsize Parallel construction process of WPPI networks}}
\label{alg401}
\begin{algorithmic}[1]
\REQUIRE ~\\
    $Path_{PPI}$: the path of PPI networks stored on HDFS;\\
    $Path_{GED}$: the path of GED datasets stored on HDFS.\\
\ENSURE ~\\
    $RDD_{WPPI}$: the RDD object of the WPPI network.
\STATE $conf$ $\leftarrow$ new SparkConf(``WPPI'', ``SparkMaster'');
\STATE $sc$ $\leftarrow$ new SparkContext($conf$);
\STATE $RDD_{PPI}$ $\leftarrow$ $sc$.textFile($Path_{PPI}$);
\STATE $RDD_{A^{W}}$ $\leftarrow$ $sc$.parallelize($RDD_{PPI}$).\textbf{map}
\STATE \quad $e$ $\leftarrow$ get edge ($RDD_{PPI}$);
\STATE \quad ($x_{i}, ~x_{j}$) $\leftarrow$ get protein-pair ($e$);
\STATE \quad $RDD_{GED}$ $\leftarrow$ loadGenes ($x_{i}$, $x_{j}$, $Path_{GED}$);
\STATE \quad $RDD_{P}$ $\leftarrow$ $RDD_{GED}$.\textbf{flatMap}
\STATE \qquad ($g_{ai},~g_{bi}$) $\leftarrow$ get genes ($RDD_{GED}$);
\STATE \qquad Pearson correlation coefficient $P_{i}$ $\leftarrow$ $Join(g_{ai}$, $g_{bi})$;
\STATE \quad \textbf{end flatMap}
\STATE \quad $RDD_{P}$.\textbf{reduceBykey}();
\STATE \quad edge weight $a^{w}$ $\leftarrow$ $RDD_{P}$.
\textbf{map}(\_$\times a^{w}$);
\STATE \textbf{end map}
\STATE $RDD_{WPPI}$ $\leftarrow$ $RDD_{A^{W}}$.\textbf{reduce}().\textbf{cache}();
\RETURN $RDD_{WPPI}$.
\end{algorithmic}
\end{algorithm}

\subsection{Parallelization of MLPCD using Spark GraphX}

\subsubsection{RDD Dependence of WPPI Network Graphic Data}
\begin{figure}[!ht]
\setlength{\abovecaptionskip}{0pt}
\setlength{\belowcaptionskip}{0pt}
\centering
\includegraphics[width=3.4in]{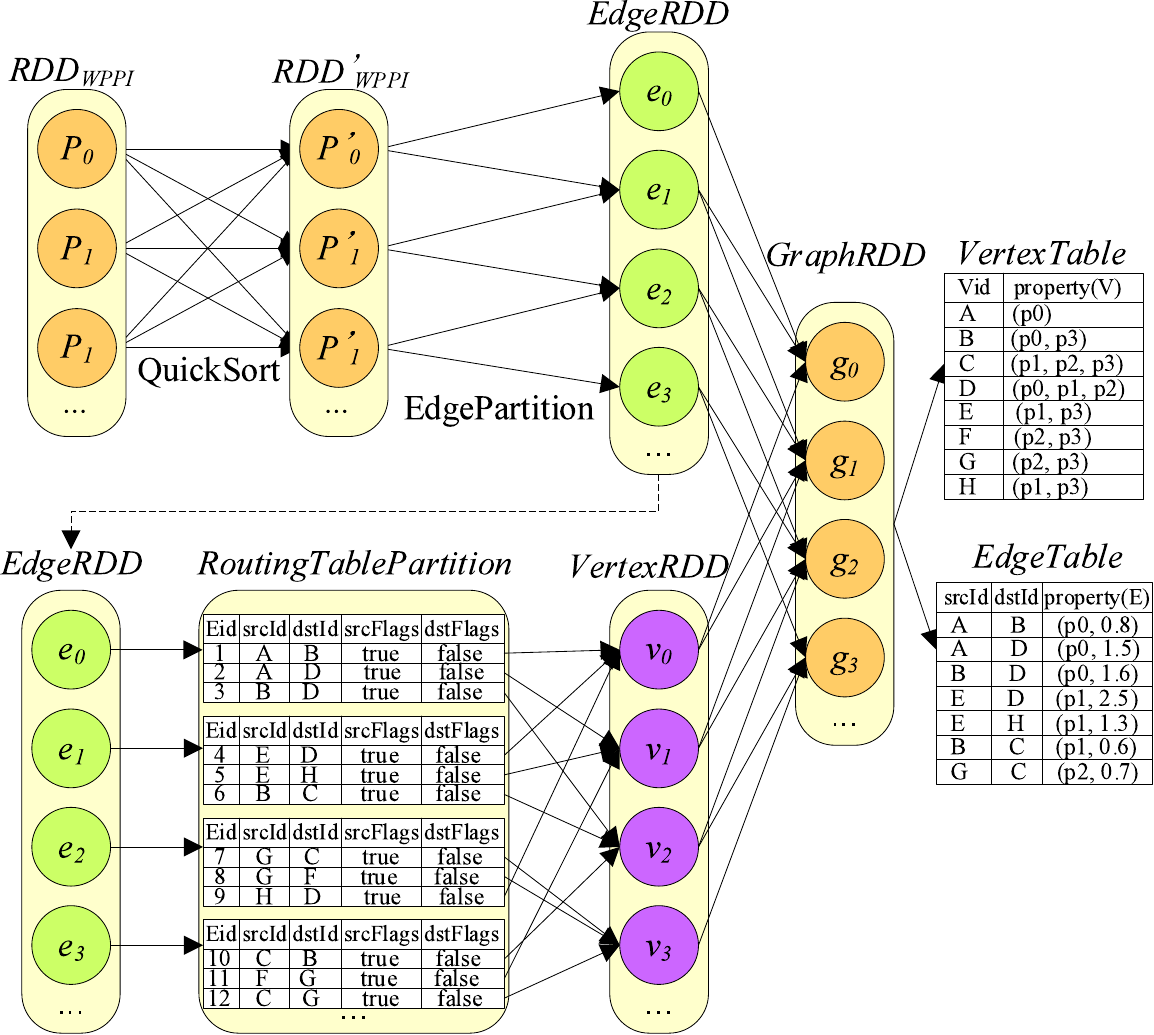}
\caption{RDD dependence of graphic data for WPPI networks.}
\label{fig04}
\end{figure}

GraphX is a parallel programming model for graph algorithms implemented on Spark, providing a rich API interface.
A graphic object of the GraphX model is the entrance of graph operations, consisting of two parts, such as edges and vertices.
All of the edges of a graph make up an EdgeRDD object, and all of the vertices make up a VertexRDD.
Then, the EdgeRDD and VertexRDD objects are joined to generate a Graph object.
The RDD dependence of a graphic WPPI network is illustrated in Fig. \ref{fig04}.

The EdgeRDD object is obtained from a $RDD_{WPPI}$, using the $quickSort()$ and $EdgePartition()$ functions.
Each edge contains a source vertex identifier ($srcId$), a destination vertex identifier ($dstId$), and other properties.
In the $quickSort()$ function, all of the records of $RDD_{WPPI}$ are re-sorted and shuffled, and new partitions are generated to form a new RDD object $RDD_{WPPI}^{'}$.
All of the edges in each partition are re-sorted by the $srcId$ in an ascending order, which can accelerate the access speed of edges.
Moreover, edges with the common vertices are allocated to the same partition as much as possible.
In the $EdgePartition()$ function, all of the edges in each partition in $RDD_{WPPI}^{'}$ are extracted, and related new partitions are established to form a new EdgeRDD object $RDD_{Edge}$.

On the basis of $RDD_{Edge}$, the VertexRDD object $RDD_{Vertex}$ is constructed according to a routing table $RoutingTablePartition$.
Distinct from the EdgeRDD, each vertex in VertexRDD is an isolated island.
To find the related edge, each vertex must save properties that the partition Id (PID) of the related edges, which are kept in the $RoutingTablePartition$.
The $RDD_{Vertex}$ and $RDD_{Edge}$ are joined to generate the graph object $RDD_{Graph}$.

\subsubsection{Protein Vertex-Cut of WPPI Network Graphic Data}
\begin{figure}[!ht]
\setlength{\abovecaptionskip}{0pt}
\setlength{\belowcaptionskip}{0pt}
\centering
\includegraphics[width=2.5in]{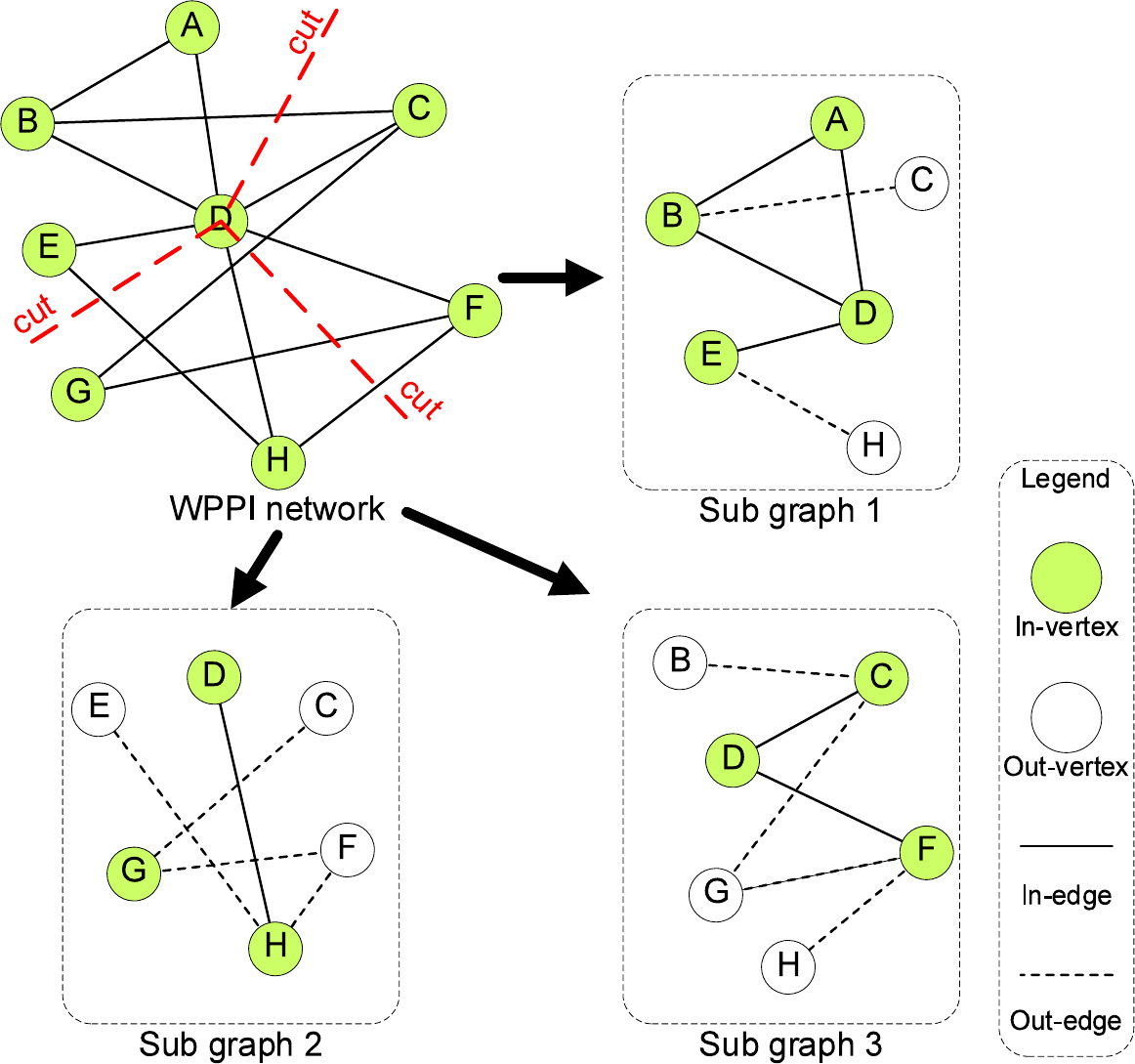}
\caption{Example of the protein vertex-cut method for WPPI networks.}
\label{fig05}
\end{figure}

The $RDD_{WPPI}$ object of WPPI network is stored in a distributed environment by a vertex-cut method using the Spark GraphX model.
In such a storage method, each edge appears in only one computing node, while each vertex might be allocated to different computers.
An example of the protein vertex-cut method for WPPI networks is illustrated in Fig. \ref{fig05}.

As shown in Fig. \ref{fig05}, the WPPI network is divided into 3 partitions by cutting the protein vertex $D$, and 3 WPPI subgraphs are generated accordingly.
Then, data of these PPI subgraphs are allocated to different computers of the Spark cluster.
All proteins are computed in parallel on different computers, which in interaction with their neighbors.

\subsubsection{Parallel Execution Process of MLPCD Algorithm}
Similar to the parallel construction process of the WPPI network, a job scheduling module $DAGScheduler$ of Spark is available for the submitted detection job.
$DAGScheduler$ analyzes the detection job and divides it into multiple job stages according to the RDD dependence of the job.

A new edge RDD object $RDD_{Edge}$ is created from $RDD_{WPPI}$ using the $Graph.fromEdgeTuples()$ function.
Then, a vertex RDD object $RDD_{Vertex}$ is generated based on $RDD_{Edge}$, and a graphic object $RDD_{Graph}$ is obtained based on them using the $Graph()$ function.
Hence, the graph object of the WPPI network is constructed in the detection process of protein communities.

After obtaining the protein subgraphs $RDD_{subG}$, the functional cohesion $FC$ of all of the subgraphs is derived.
If the value of $FC$ in the current splitting scenario is bigger than that of $FC_{Best}$, namely, the current splitting scenario is the best splitting scenario.
All of the detected protein subgraphs are appended to the RDD object $RDD_{WPPI}$ as the final protein communities.
Otherwise, each one of the current subgraphs continues being divided iteratively.
The detailed steps of the parallel MLPCD algorithm based on the Spark GraphX model are presented in Algorithm \ref{alg402}.

\begin{algorithm}[!ht]
\scriptsize
\caption{{\scriptsize Parallel MLPCD algorithm}}
\label{alg402}
\begin{algorithmic}[1]
\REQUIRE ~\\
    $RDD_{WPPI}$: the RDD object of a WPPI network;\\
    $d_{\alpha}$: the threshold of the initial hub vertices; \\
    $\lambda$: the threshold of functional cohesion.
\ENSURE ~\\
    $RDD_{PCs}$: the RDD object of the detected protein communities.
\STATE $conf$ $\leftarrow$ new SparkConf (``MLPCD'', ``SparkMaster'');
\STATE $sc$ $\leftarrow$ new SparkContext($conf$);
\STATE $RDD_{Cs} \leftarrow$ $sc$.parallelize($RDD_{WPPI}$).\textbf{mapVertices}
\STATE \quad each vertex $v_{i}$:
\STATE \quad calculate the weighted degree $d(v_{i})$;
\STATE \quad \textbf{if} {$d(v_{i})$ $>$ $d_{\alpha}$} \textbf{then}
\STATE \qquad select $v_{i}$ as initial community $\rightarrow Cs$;
\STATE \textbf{endmap}.collect().reduce();
\STATE $RDD_{Pcs} \leftarrow$ $RDD_{PCs}$.parallize().\textbf{flatMap}
\STATE \quad each community $C_{k}$:
\STATE \quad collect neighbors $N(C_{k})$ of vertices in $C_{k}$;
\STATE \quad  $\Delta{Q} \leftarrow$ $N(C_{k})$.\textbf{foreach}
\STATE \qquad  each vertex $v_{j}$:
\STATE \qquad calculate the modularity $\Delta{Q_{C_{k},v_{j}}} \leftarrow Q_{C_{k} \cup v_{j}} - Q_{C_{k}}$;
\STATE \quad \textbf{endfor}.shuffle().reduce();
\STATE \quad append the neighbor with max $\Delta{Q}$ to $C_{k}$;
\STATE \quad \textbf{endmap}.groupByKey().reduce();
\STATE reconstruct new network $RDD_{G'}$ based on communities $RDD_{PCs}$;
\STATE initial vertices as communities;
\STATE $sc$.parallelize($RDD_{G'}$).\textbf{mapVertices}
\STATE \quad each vertex $v_{i}'$:
\STATE \quad collect neighbors $N(v_{i}')$ of $v_{i}'$;
\STATE \quad $FC_{Best} \leftarrow$ $N(v_{i}')$.parallelize().\textbf{flatMap}
\STATE \qquad each vertex $v_{j}'$:
\STATE \qquad append $v_{j}'$ to the community $C_{k}'$ that $v_{i}'$ located;
\STATE \qquad calculate functional cohesion $FC_{k} \leftarrow II_{k} \times Con_{k}$;
\STATE \qquad \textbf{if} {$FC_{k} < \lambda$} \textbf{then}
\STATE \quad \qquad remove $v_{j}'$ from the community $C_{k}'$;
\STATE \quad \textbf{endmap}.reduce();
\STATE \textbf{endmap}.groupByKey();
\RETURN $RDD_{PCs}$.
\end{algorithmic}
\end{algorithm}

The whole process of the parallel MLPCD algorithm includes the parallel construction process of the WPPI network and the protein community parallel detection of the WPPI network.
The time complexity of the parallel construction process of WPPI network is $O(\frac{mM}{p})$, where $m$ is the number of edges in the WPPI network, $M$ is the number of samples of each GED record, and $p$ is the number of computers in the Spark cluster.
The time complexity of the protein community parallel detection is $O(\frac{NK_{1}}{\log p})$.
Therefore, the time complexity of the whole process of the MLPCD algorithm is $O(\frac{mM}{p}+\frac{NK_{1}}{\log p})$.

\section{Experiments and Applications}
\label{section5}

\subsection{Experimental Setup}
\label{section5.1}
All the experiments are conducted on an Apache Spark cloud platform at the National Supercomputing Center in Changsha \cite{a27}.
Each computer node uses Ubuntu 12.04.4 and has one Pentium (R) Dual-Core 3.20GHz CPU and 32GB memory.
The original PPI datasets are downloaded from the Database of Interacting Proteins (DIP) \cite{a28} and STRING \cite{a29} databases.
The related gene expression datasets are collected from the Gene Expression Omnibus (GEO) of NCBI \cite{a30} and Array Express (AE) of EBI \cite{a31}.
The PPI and GED datasets used in the experiments are provided in Table \ref{table51}.

\begin{table}[!ht]
\scriptsize
\renewcommand{\arraystretch}{1.0}
\setlength{\abovecaptionskip}{0pt}
\setlength{\belowcaptionskip}{0pt}
\caption{{\scriptsize PPI and GED datasets used in the experiments.}}
\label{table51}
\centering
\tabcolsep1pt
\begin{tabular}{L{1.7cm} C{1.3cm} C{1.5cm} C{1.4cm} C{0.8cm} C{1.5cm}}
\hline
PPI datasets & \#.Proteins & \#.Interactions &GED datasets &\#.Genes &Matching ratio\\
\hline
A. laidlawii  &6,373 &85,733  &GDS3823 &22,810  & 96.43\%\\
C. hominis    &6,645 &92,373  &GDS976 &5,760  &84.50\%\\
D. africanus  &3,679 &347,162 &GDS290 &5,355 &85.71\%\\
E. bacterium  &4,118 &426,093 &GDS2353 &15,720 &89.34\%\\
B. clausii    &4,062 &428,565 &GDS2181 &4,290  &82.56\%\\
E. coli       &4,590 &505,207 &GDS3597 &10,208 &81.85\%\\
F. johnsoniae &4,977 &574,383 &GDS2388 &18,968 &91.21\%\\
S. cerevisiae &6,391 &1,003,567 &GDS2505 &10,807  &98.27\%\\
H. sapiens    &19,576 &5,676,528 &GDS4798 &54,675 &85.41\%\\
M. musculus   &21,151 &6,307,021 &GDS3462 &45,101 &86.90\%\\
\hline
\end{tabular}
\end{table}

\subsection{Experimental Results Analysis}
\label{section5.2}

\begin{table*}[!pt]
\scriptsize
\renewcommand{\arraystretch}{1.0}
\setlength{\abovecaptionskip}{0pt}
\setlength{\belowcaptionskip}{0pt}
\caption{{\scriptsize Examples of matched protein complexes of \emph{S. cerevisiae} detected by MLPCD.}}
\label{table52}
\centering
\tabcolsep1pt
\begin{tabular}{L{2.7cm} C{7.0cm} C{1.5cm} C{1.0cm} C{1.0cm} C{1.0cm} C{1.0cm} C{1.0cm} C{1.0cm}}
\hline
\multicolumn{2}{c}{} & Known & \multicolumn{3}{c}{Overlap values} & \multicolumn{3}{c}{Overlap values}\\
\multicolumn{2}{c}{Matched protein complexes} & complexes & \multicolumn{3}{c}{on PPI network} & \multicolumn{3}{c}{on WPPI network}\\
\hline
Complex name & Proteins &Size &Size &Overlap &OS &Size &Overlap &OS \\
\hline
Swr1p complex &	YNG2, EAF1, ESA1, EPL1, EAF3, TRA1 & 6 &3 &3 &50.00\% &7 &6 &100.00\% \\
Cdc3 complex & CDC11, CDC12, CDC3, CDC10, YDL225W &8 &6 &3 &37.50\% &8 &5 &62.50\% \\
Prefoldin complex &	GIM4, PAC10, YKE2, TUB4, GIM5, GIM3 &8 &12 &3 &37.50\% &10 &6 &75.00\% \\
Lsm complex &LSM7, LSM6, LSM1, LSM2, LSM3, LSM4, LSM5, LSM8, DCP1, PAT1, KEM1 & 14 & 16 & 8 & 57.14\% & 14 & 11 & 78.57\% \\
NuA4 histone acetyltransferase complex & RVB1, VPS71, VPS72, SWR1, SWC3, SWC5, RVB2, ACT1, YAF9, SWC4, ARP4 &16 &9 &7 & 43.75\% &13 &11 &68.75\% \\
SAGA complex &SPT7, HFI1, SPT20, TAF61, SPT15, SPT8, TAF25, SPT3, TRA1, TAF60, TAF90, GCN5, TAF17, NGG1, ADA2 &20 &7 & 4 &20.00\% &22 &15 &75.00\% \\
\hline
\end{tabular}
\end{table*}
\begin{figure*}[!pt]
\setlength{\abovecaptionskip}{0pt}
\setlength{\belowcaptionskip}{0pt}
\centering
 \includegraphics[width=7.0in]{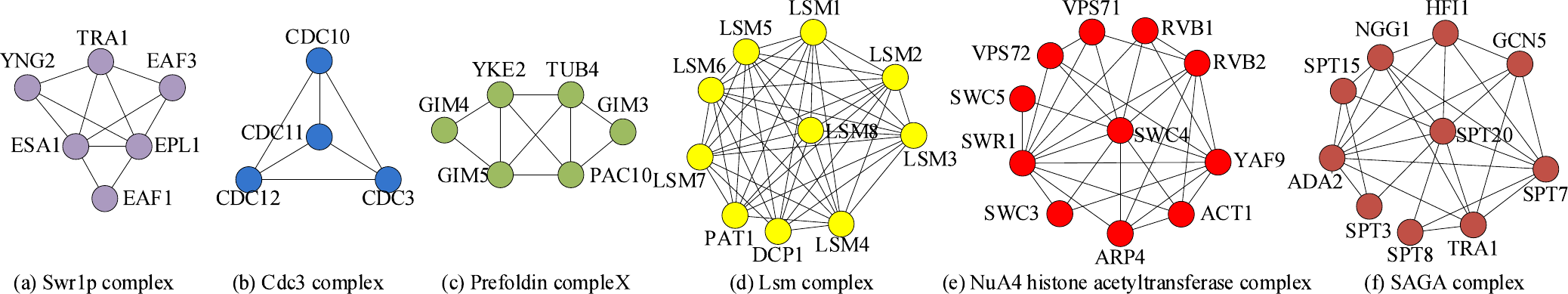}
 \caption{Examples of protein complexes of \emph{S. cerevisiae} detected by MLPCD.}
 \label{chart01}
\end{figure*}

\subsubsection{Protein Complexes Analysis}
The MLPCD algorithm is applied to the 10 groups of PPI datasets described in Table \ref{table51}.
We compare the identified protein communities with the known protein complexes using the protein complex catalogue in the MIPS \cite{a32} database.
We introduce the overlapping score to evaluate the efficacy of the identified protein communities matching to known protein complexes.
The overlapping score $OS(Pc, Kc)$ between an identified protein community $Pc$ and a known protein complex $Kc$ is defined in Eq. (\ref{eq51}):

\begin{equation}
\label{eq51}
OS(Pc, Kc) = \frac{|V_{Pc} \cap V_{Kc}|^{2}}{|V_{Pc}| \times |V_{Kc}|},
\end{equation}
where $|V_{Pc}|$ and $|V_{Kc}|$ are the number of proteins in $Pc$ and $Kc$, respectively.
$|V_{Pc} \cap V_{Kc}|$ is the number of proteins exist in both $Pc$ and $Kc$.
If the value of $OS(Pc, Kc)$ is larger than a specific threshold, then the identified $Pc$ is considered as matching to $Kc$.
Taking \emph{S. cerevisiae} as an example, where there are 532 known protein complexes published in MIPS, 405 out of 763 protein communities detected by MLPCD are considered to be matched with the known protein complexes.
Due to the incompleteness of the known complexes in MIPS, the non-matched communities might provide potential candidate complexes for biologists to further validate.
Examples of the matched protein complexes of \emph{S. cerevisiae} are shown in Table \ref{table52}.

In Table \ref{table52}, six examples are given to show how MLPCD detect protein communities more accurately on WPPI networks than that on the original PPI networks.
Take the \emph{Lsm complex} as an example, 11 proteins are matched in a 14-member community detected on the WPPI network ($OS = 78.57\%$), while only 8 proteins are matched in a 16-member community detected on the original PPI network ($OS =57.14\%$).
For the \emph{SAGA complex}, 15 proteins are matched in a 22-member community detected on the WPPI network ($OS = 75.00\%$), while only 4 proteins are overlapped in a 7-member community detected on the original PPI network ($OS =20.00\%$).
These results indicate that interaction relationship alone is not enough to identify protein complexes, and the integration of the GED is useful for detection of protein communities with more biological significance.
Parts of the detected protein complexes of the \emph{S. cerevisiae} PPI network are shown in Fig. \ref{chart01}.

\begin{figure}[!ht]
\setlength{\abovecaptionskip}{0pt}
\setlength{\belowcaptionskip}{0pt}
\centering
 \includegraphics[width=2.6in]{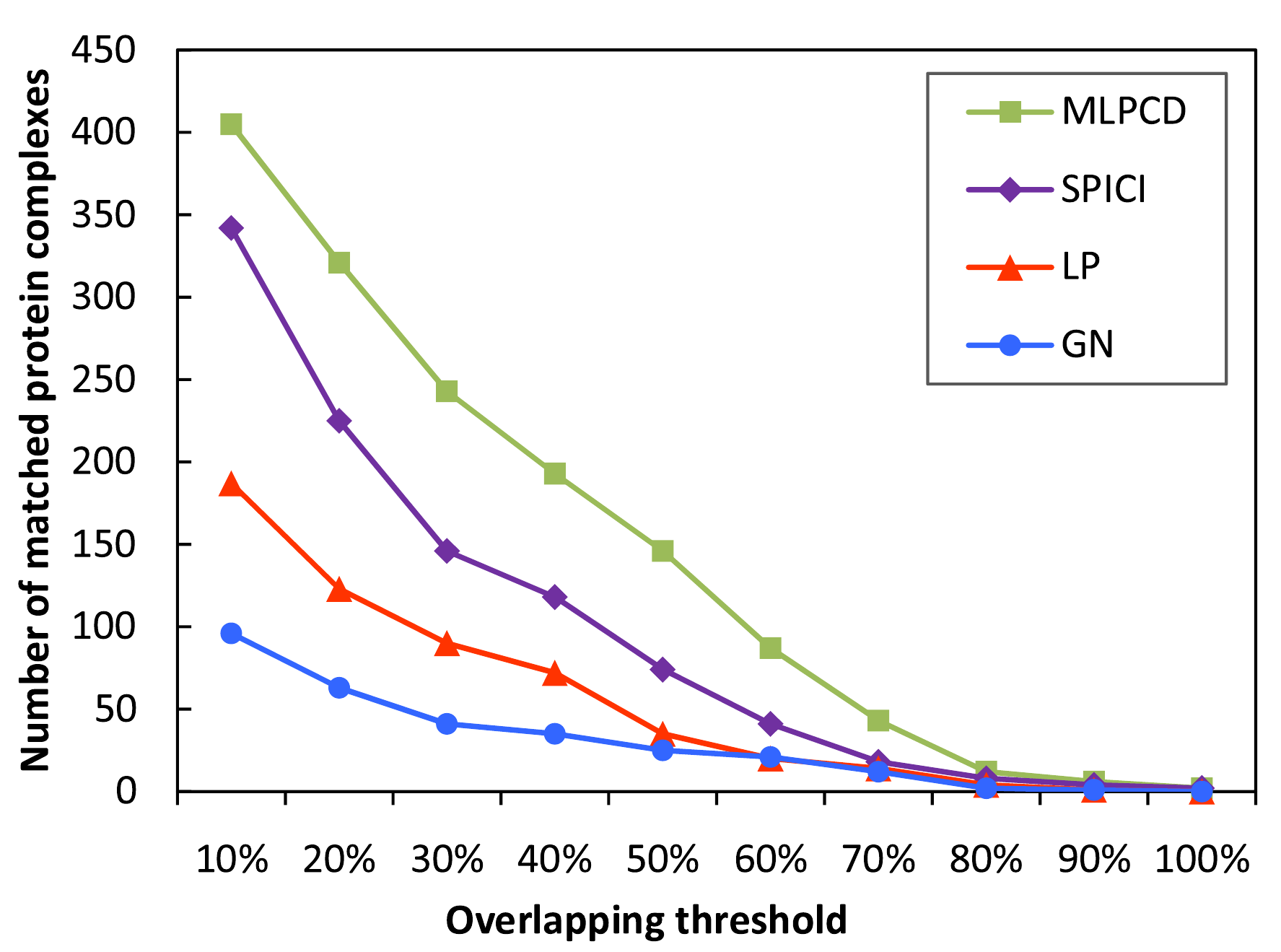}
 \caption{Protein complexes detection comparison.}
 \label{chart02}
\end{figure}

We further evaluate the preciseness of MLPCD by comparing with SPICI \cite{a07}, LP \cite{a06}, and GN \cite{a04} algorithms on \emph{S. cerevisiae}.
The comparison results are shown in Fig. \ref{chart02}.

As the overlapping threshold increases, the matching condition of protein complexes becomes more stringent.
In such a way, the number of matched protein complexes in each comparison algorithm decreases obviously.
For example, when the overlapping threshold is equal to 10.00\%, there are 405 matched complexes that identified by MLPCD,  342 matched complexes for SPICI,  and 187 matched complexes for LP.
However, when the overlapping threshold rises to 60.00\%, there are only 87 matched complexes that identified by MLPCD,  41 matched complexes for SPICI,  and 20 matched complexes for LP.
In addition, the comparison results in Fig. \ref{chart02} indicate that MLPCD can identify more protein complexes than other algorithms in each case of the same overlapping threshold.

\begin{table*}[!pt]
\scriptsize
\renewcommand{\arraystretch}{1.0}
\setlength{\abovecaptionskip}{0pt}
\setlength{\belowcaptionskip}{0pt}
\caption{{\scriptsize Examples of matched protein functional modules of \emph{S. cerevisiae} detected by MLPCD.}}
\label{table53}
\centering
\tabcolsep1pt
\begin{tabular}{L{5.0cm} C{11.0cm} C{1.5cm}}
\hline
Biological processes & Proteins in functional modules & P-value \\
\hline
Phospholipid metabolic process	& TVP38, MRL1, DPP1, YNL194C & 6.45E-03 \\
ER-associated protein catabolic process	& HRD3, HRD1, HMG2 &  1.62E-04 \\
Deoxycytidine catabolic process	& DBP10, MRT4, RPF1, TIF6, NOP15, BRX1 & 6.06E-04\\
Regulation of carbohydrate metabolic process & MFT1, NBP2, SSK2, YRA2, MEX67, PTC1, HAP5, UTP13, YRA1, QRI8, PRP11, SLX5, SSK1, HAP2, PBS2, MTH1, SUB2, HAP3, GBP2  & 2.79E-07 \\
Ribonucleoside monophosphate biosynthetic process & PRS4, IME1, TPS2, TPS3, PRS2, ARG80, PRS3, ARG81, NBA1, MCM1, PRS5, KAP114, TPS1, TSL1, PRS1, NIS1, RIM11, UME6, NAP1 & 4.61E-12 \\
DNA replication process	& PRI1, RFC4, RFC5, POL1, POL30, CTF8, RAD57, RFC3, POL12, RFC1, YGL081W, RAD24, ELG1, RFC2 &  2.62E-14
\\
\hline
\end{tabular}
\end{table*}
\begin{figure*}[!ht]
\setlength{\abovecaptionskip}{0pt}
\setlength{\belowcaptionskip}{0pt}
\centering
 \includegraphics[width=7.0in]{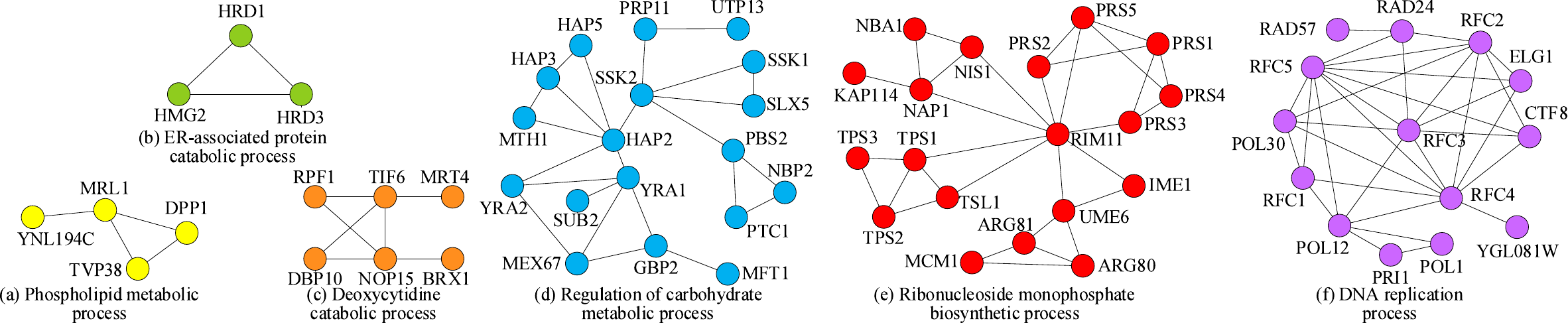}
 \caption{Examples of protein functional modules of \emph{S. cerevisiae} detected by MLPCD.}
 \label{chart03}
\end{figure*}

\subsubsection{Protein Functional Modules Analysis}
We use Gene Ontology (GO) annotations to evaluate the functional enrichment of the identified protein communities, which are downloaded from Gene Ontology Consortium \cite{a33}.
Modules of the detected protein communities are statistically evaluated using the $P$-value \cite{a34} from the hypergeometric distribution, which is defined in Eq. (\ref{eq52}):

\begin{equation}
\label{eq52}
P = 1 - \sum_{i=0}^{k-1}{\frac{C_{|F|}^{i} \times C_{|V|-|F|}^{|C|-i}}{C_{|V|}^{|C|}}},
\end{equation}
where $|V|$ is the total number of proteins in a WPPI network, $|C|$ is the size of the identified communities, $|F|$ is the number of a known protein functional group of biological processes annotated by Gene Ontology (GO), and $k$ is the number of proteins in common between the protein functional module and the identified community.
$P$-value is also known as a metric of functional homogeneity, which is the probability that at least $k$ proteins in a module of size $|C|$ are included in a functional group of size $|F|$.
A low $P$-value indicates that the module closely corresponds to the protein communities, because it is less probable that the network will produce the module by chance.
The smaller $P$-value implies little randomness of protein community.
Consequently, the protein community with a $P$-value below the minimum threshold has higher biological significance than the protein community with a high $P$-value.
Examples of the matched protein functional modules  of \emph{S. cerevisiae} are shown in Table \ref{table53} and Fig. \ref{chart03}.

As shown in Table \ref{table53} and Fig. \ref{chart03}, proteins of \emph{DBP10}, \emph{MRT4}, \emph{RPF1}, \emph{TIF6}, \emph{NOP15}, \emph{BRX1} in the same module are related to the biological process of \emph{Deoxycytidine catabolic process}.
Proteins of \emph{HRD3}, \emph{HRD1}, \emph{HMG2} in the same module are related to the \emph{ER-associated protein catabolic process}.
In addition, the unknown function protein \emph{YNL194C} is included in a 4-member community, of which the other three proteins are related to the \emph{Phospholipid metabolic process}.
Therefore, we can predict that \emph{YNL194C} is also a \emph{Phospholipid metabolic} functional protein.
The unknown function protein \emph{YGL081W} is included in a 14-member community, of which the other three proteins are related to the process of \emph{DNA replication}.
Therefore, we can predict that \emph{YGL081W} is also a \emph{DNA replication} protein.
Experimental results indicate that most of the detected protein communities are enriched for proteins with the same or similar biological processes.

\subsubsection{ROC Curve Analysis for Different Algorithms}
We compare the experimental results of our proposed MLPCD algorithm and the SPICI, LP, and GN algorithms by analyzing the results of protein communities detection.
Due to space limitation, the results of the experiment are provided in the supplementary file.

\subsection{Performance Evaluation}

\subsubsection{Execution Time for Different PPI Networks}
\begin{figure}[!ht]
\setlength{\abovecaptionskip}{0pt}
\setlength{\belowcaptionskip}{0pt}
\centering
\includegraphics[width=2.5in]{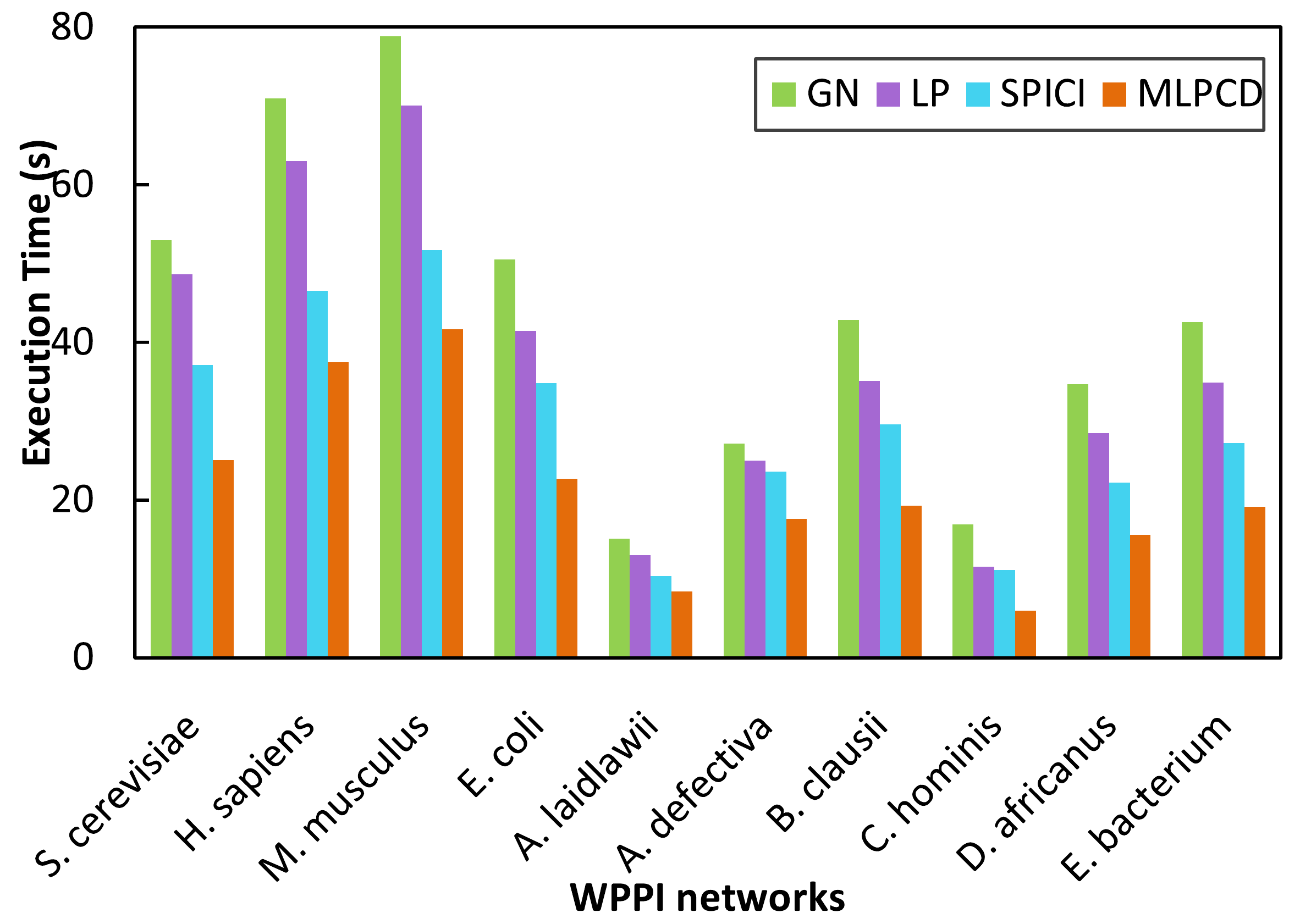}
\caption{Execution time of the comparison algorithms.}
\label{chart05}
\end{figure}

Experiments are performed to compare the performance of MLPCD with that of the SPICI, LP, and GN algorithms.
10 groups of PPI network datasets in Table \ref{table51} are used in the experiments.
In these experiments, the number of computers in the Spark cloud environment is set to 5.
For each comparison algorithm, the execution time of protein community detection from WPPI networks is recorded and compared.
The experimental results are presented in Fig. \ref{chart05}.

It is clear from Fig. \ref{chart05} that the proposed MLPCD algorithm outperforms others on each dataset.
For example, in the \emph{D. africanus} case, there are 3,679 proteins and 347,162 interactions, the execution time of MLPCD is approximate to 15.62s, while that of SPICI is 22.21s, that of LP is 28.46s, and that of GN is 34.71s.
When the scale of proteins in a PPI network is greater than 10, 000, the performance advantage of algorithm MLPCD  is even more obvious.
For example, in the \emph{M. musculus} case, there are 21,151 proteins and 6,307,021 interactions, the execution time of MLPCD is approximate to 41.62s, while that of SPICI is 51.72s, that of LP is 70.01s, and that of GN is 78.84s.
Taking advantage of the task parallel optimization, MLPCD achieves significant strengths over SPICI, LP, and GN algorithms in terms of performance.

\subsubsection{Speedup Evaluation of MLPCD}
\begin{figure}[!ht]
\setlength{\abovecaptionskip}{0pt}
\setlength{\belowcaptionskip}{0pt}
\centering
\includegraphics[width=3.0in]{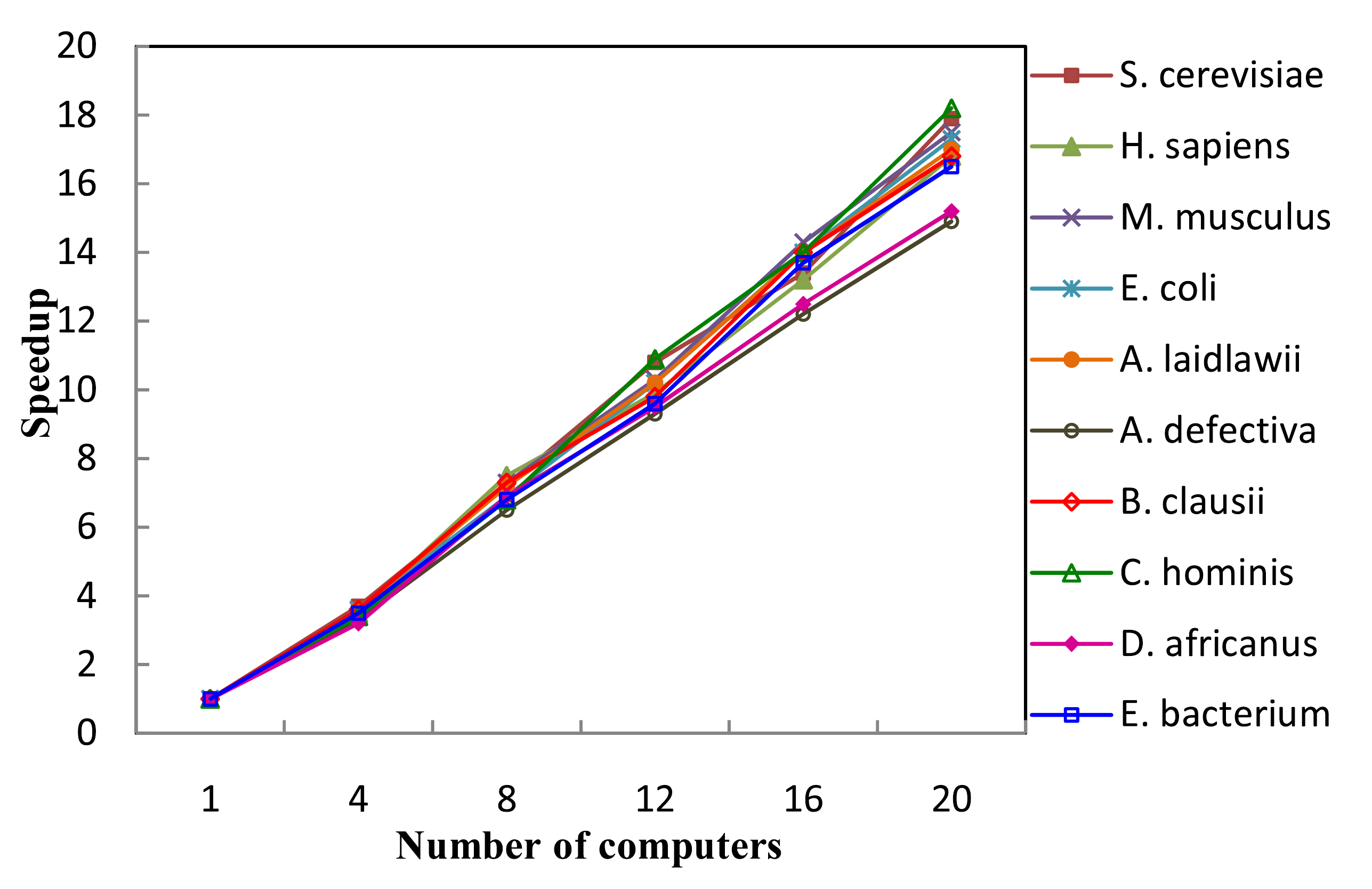}
\caption{Speedup of MLPCD for different computer scales.}
\label{chart06}
\end{figure}

The speedup of MLPCD is evaluated in a Spark computing cluster, where the number of computers gradually increases from 1 to 20.
The average execution time of MLPCD on 10 PPI datasets is described in Table \ref{table51} is recorded.
The experimental results are presented in Fig. \ref{chart06}.

As shown in Fig. \ref{chart06}, benefiting from the parallel optimization, the speedup of MLPCD tends to increase in each experiment with the number of computers increases.
When the number of computers increases from 4 to 20, the speedup of MLPCD in each case shows a rapid growth trend.
The average speedup of MLPCD in all cases is 3.48 on 4 computers and 7.04 on 8 computers, respectively.
When the number of computers is equal to 20, the speedup of MLPCD in all cases is in the range of 14.9 - 18.2.

\section*{Conclusions}
In this paper, we proposed a MLPCD algorithm to detect protein communities from large-scale PPI networks.
By integrating GED datasets in specific conditions, we constructed a WPPI network based on gene co-expression of proteins in the original PPI network.
Based on the defined community modularity and functional cohesion, protein communities are detected from WPPI in an agglomerative way.
We evaluated the identified protein communities with known protein complexes and evaluate the functional enrichment of protein functional modules using Gene Ontology annotations.
Experimental results indicated the superiority and notable advantages of the MLPCD algorithm over the relevant algorithms in terms of accuracy and efficiency.

For future work, we will further concentrate on the knowledge discovery of PPI networks and related research fields.
The research on dynamic characteristics of PPI networks is one of the most challenging issues in the field of Bioinformatics.

\section*{Acknowledgment}
The research was partially funded by the Key Program of National Natural Science Foundation of China (Grant No. 61432005),
the National Outstanding Youth Science Program of National Natural Science Foundation of China (Grant No. 61625202),
and the National Natural Science Foundation of China (Grant No. 61370095, 61772182).
This work is also supported by a UIC Chancellor's Graduate Research Fellowship and UIC CCTS Pre-doctoral Education for Clinical and Translational Scientists fellowship (UL1TR002003).

\ifCLASSOPTIONcaptionsoff
\newpage
\fi

\bibliographystyle{IEEEtran}
\bibliography{reference}

\begin{IEEEbiography}
[{\includegraphics[width=1in, height=1.25in, clip, keepaspectratio]{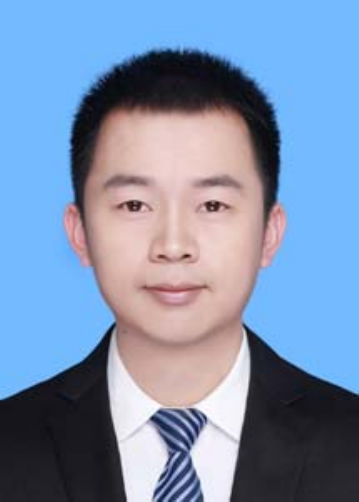}}]
{Jianguo Chen}  received the Ph.D. degree in College of Computer Science and Electronic Engineering at Hunan University, China.
He was a visiting Ph.D. student at the University of Illinois at Chicago from 2017 to 2018.
He is currently a postdoctoral in University of Toronto and Hunan University.
His major research areas include parallel computing, cloud computing, machine learning, data mining, bioinformatics and big data.
He has published research articles in international conference and journals of data-mining algorithms and parallel computing, such as
{\em IEEE TPDS}, IEEE/ACM TCBB, and {\em Information Sciences}.
\end{IEEEbiography}

\begin{IEEEbiography}
[{\includegraphics[width=1in, height=1.25in, clip, keepaspectratio]{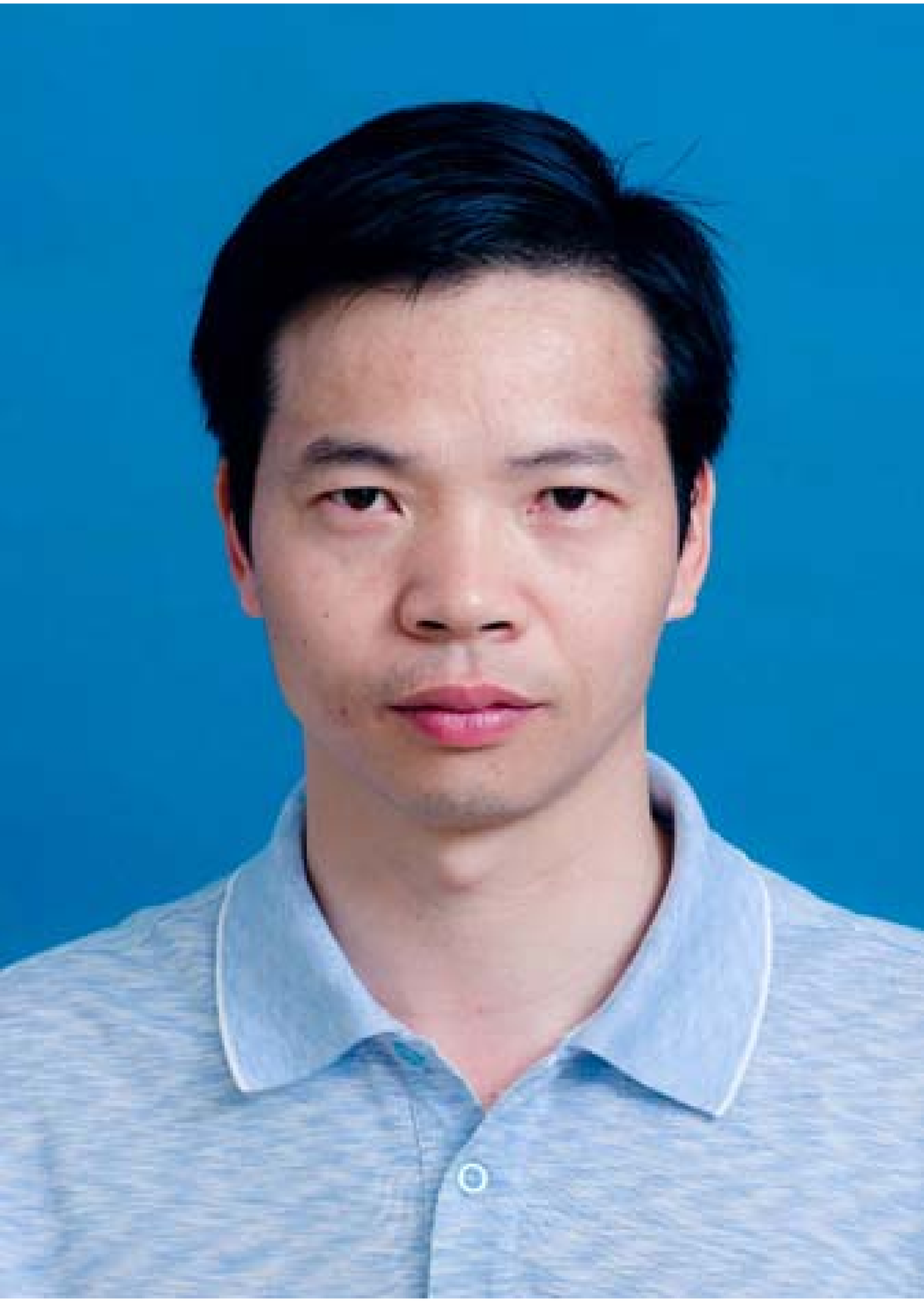}}]
{Kenli Li} received the Ph.D. degree in computer science from Huazhong University of Science and Technology, China, in 2003.
He is currently a full professor of computer science and technology at Hunan University
and director of National Supercomputing Center in Changsha.
His major research areas include parallel computing, high-performance computing, grid and cloud computing.
He has published more than 130 research papers in international conferences and journals.
He is an outstanding member of CCF and a senior member of the IEEE.
\end{IEEEbiography}

\begin{IEEEbiography}
[{\includegraphics[width=1in, height=1.25in, clip, keepaspectratio]{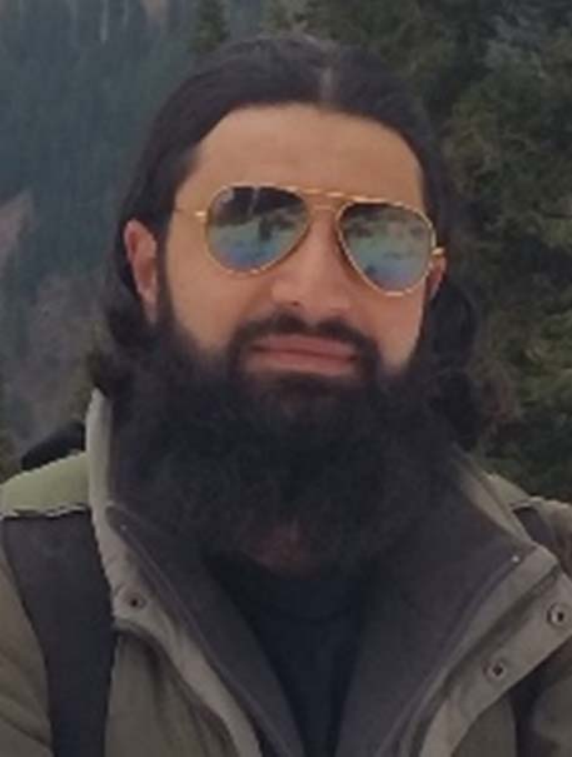}}]
{Kashif Bilal} received his PhD from North Dakota State University USA. He is currently a post-doctoral researcher at Qatar University, Qatar. His research interests include cloud computing, energy efficient high speed networks, and robustness.Kashif is awarded CoE Student Researcher of the year 2014 based on his research contributions during his doctoral studies at North Dakota State University.
\end{IEEEbiography}

\begin{IEEEbiography}
[{\includegraphics[width=1in, height=1.25in, clip, keepaspectratio]{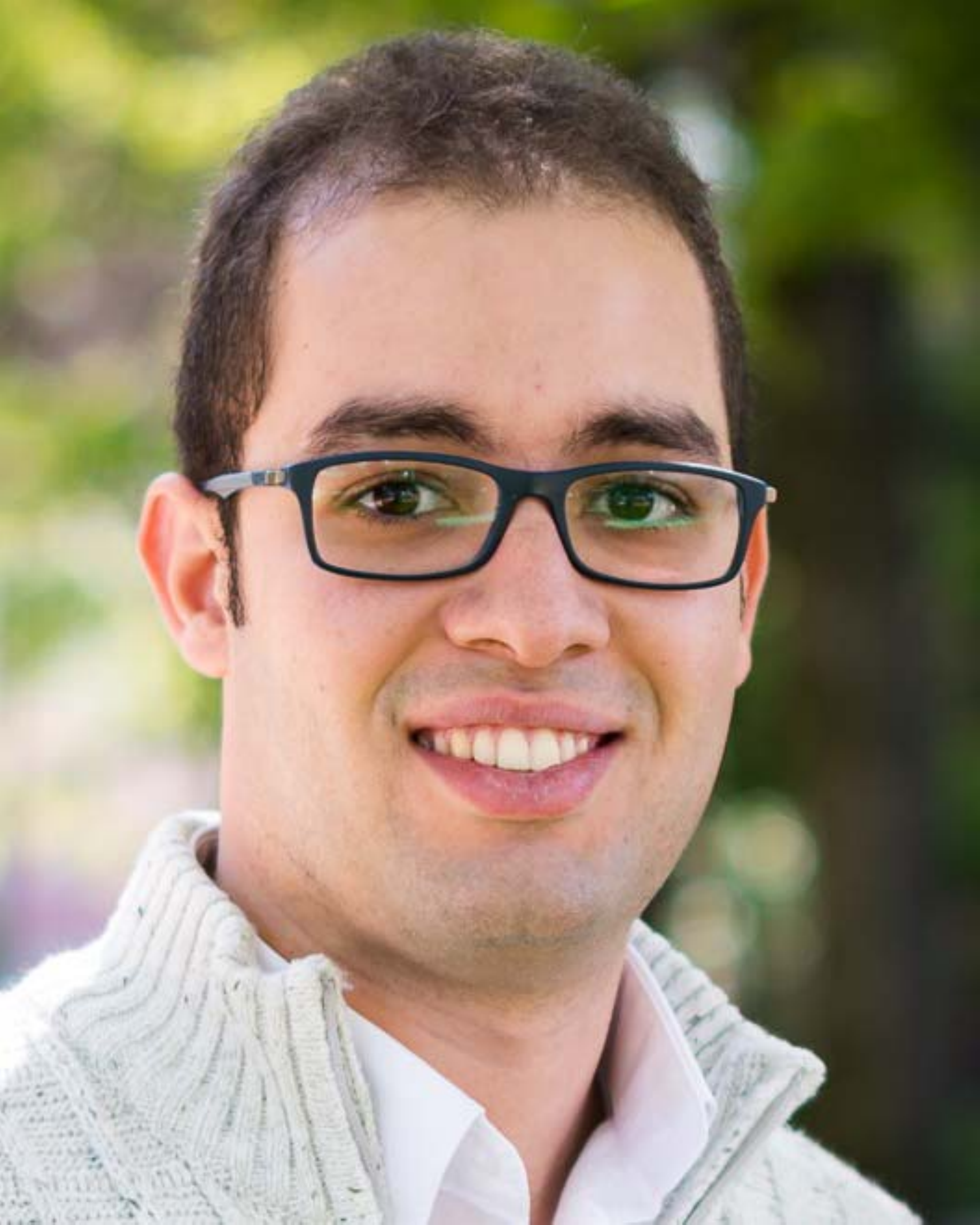}}]
{Ahmed A. Metwally} is a Bioinformatics Ph.D. candidate and CS MS candidate at the University of Illinois at Chicago.
He got his B.Sc. in 2010 with the top-class honor, and M.Sc. in 2014, both in Biomedical Engineering from Cairo University. His Ph.D. proposal received the UIC Chancellor¡¯s research award, first-place award at the UIC Research Forum, the Scientific Excellence Award at the UIC Department of Medicine scholarly activities, and 2nd place award of the ISCB GLBIO'17 conference.
\end{IEEEbiography}

\begin{IEEEbiography}
[{\includegraphics[width=1in, height=1.25in, clip, keepaspectratio]{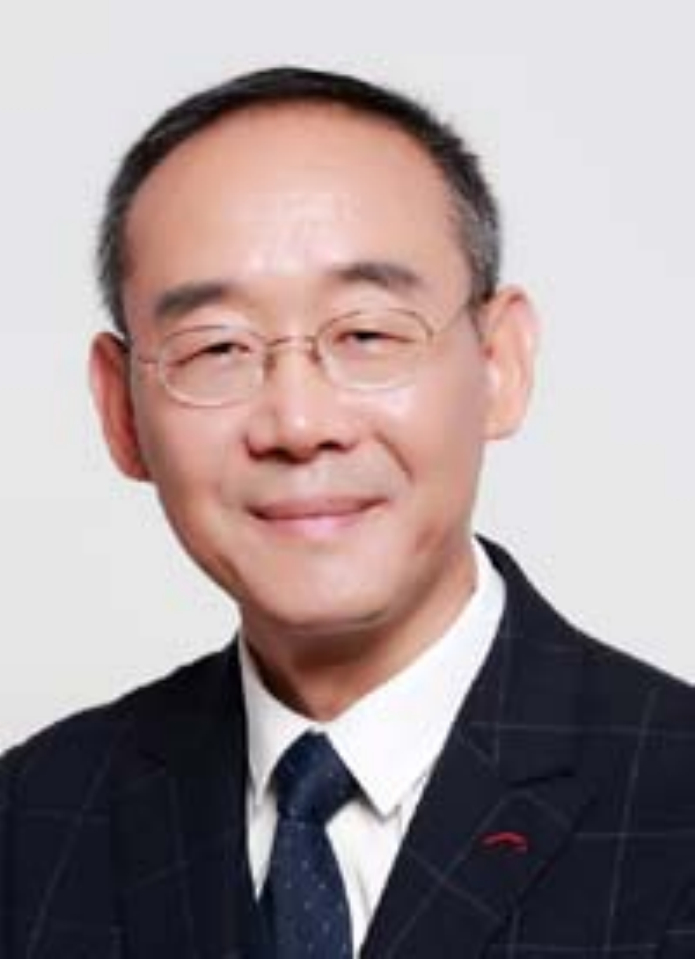}}]
{Keqin Li} is a SUNY Distinguished Professor of computer science in the State University of New York. His current research interests include parallel computing and high-performance computing, distributed computing, energy-efficient computing and communication, heterogeneous computing systems, cloud computing, big data computing, CPU-GPU hybrid and cooperative computing, multi-core computing, storage and file systems,
wireless communication networks, sensor networks, peer-to-peer file sharing systems, mobile computing, service computing, Internet of things and cyber-physical systems. He has
published over 590 journal articles, book chapters, and refereed conference papers, and
has received several best paper awards. He is currently serving or has served on the editorial boards of IEEE Transactions on Parallel and Distributed Systems, IEEE Transactions
on Computers, IEEE Transactions on Cloud Computing, IEEE Transactions on Services
Computing, and IEEE Transactions on Sustainable Computing. He is an IEEE Fellow.
\end{IEEEbiography}

\begin{IEEEbiography}
[{\includegraphics[width=1in, height=1.25in, clip, keepaspectratio]{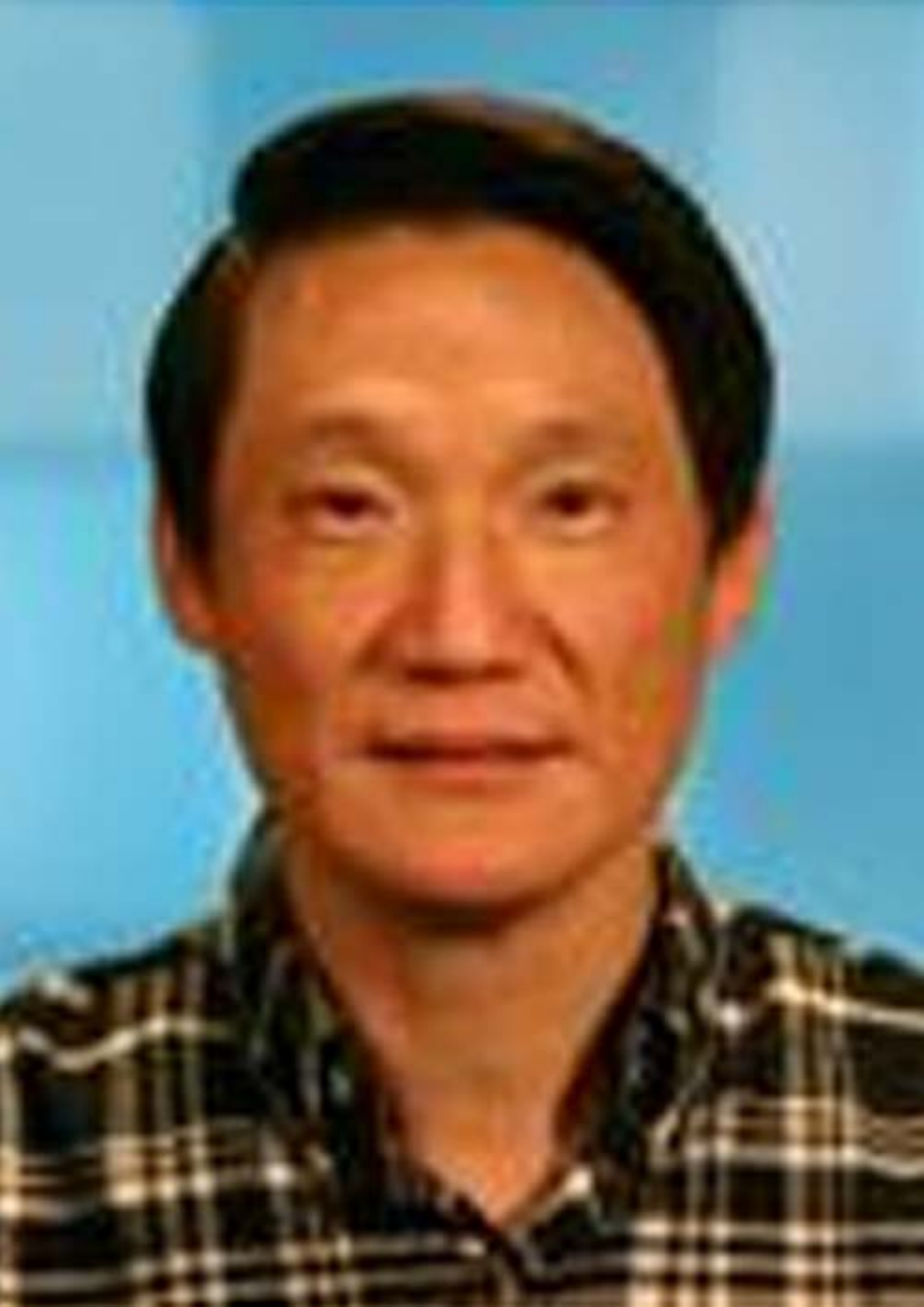}}]
{Philip S. Yu} received the B.S. Degree in E.E. from National Taiwan University, the M.S. and Ph.D. degrees in E.E. from Stanford University, and the M.B.A. degree from New York University.
He is a Distinguished Professor in Computer Science at the University of Illinois at Chicago and also holds the Wexler Chair in Information Technology.
Before joining UIC, Dr. Yu was with IBM, where he was manager of the Software Tools and Techniques department at the Watson Research Center.
His research interest is on big data, including data mining, data stream, database and privacy.
He has published more than 1,000 papers in refereed journals and conferences. He holds or has applied for more than 300 US patents.
Dr. Yu is a Fellow of the ACM and the IEEE.
He is the Editor-in-Chief of {\em ACM Transactions on Knowledge Discovery from Data}.
Dr. Yu is the recipient of ACM SIGKDD 2016 Innovation Award for his influential research and scientific contributions on mining, fusion and anonymization of big data, the IEEE Computer Society¡¯s 2013 Technical Achievement Award for ``pioneering and fundamentally innovative contributions to the scalable indexing, querying, searching, mining and anonymization of big data'', and the Research Contributions Award from IEEE Intl. Conference on Data Mining (ICDM) in 2003 for his pioneering contributions to the field of data mining.
He also received the ICDM 2013 10-year Highest-Impact Paper Award, and the EDBT Test of Time Award (2014).
\end{IEEEbiography}

\end{document}